\documentclass[11pt]{article}
\usepackage{a4}
\usepackage{amsmath}
\usepackage{amsfonts}
\usepackage{latexsym}
\usepackage{epsfig}
\renewcommand{\d}{\mathrm{d}}

\newcommand{\Ra}{\Rightarrow}

\DeclareMathSymbol{\mg}{\mathrel}{symbols}{"1D}

\newcommand{\ga}{\alpha}
\newcommand{\gb}{\beta}
\renewcommand{\gg}{\gamma}
\newcommand{\gd}{\delta}
\renewcommand{\ge}{\epsilon}

\newcommand{\gf}{\phi}

\newcommand{\gx}{\xi}
\newcommand{\gm}{\mu}
\newcommand{\gn}{\nu}

\newcommand{\gl}{\lambda}
\newcommand{\gr}{\rho}
\newcommand{\gth}{\theta}
\newcommand{\gs}{\sigma}

\newcommand{\go}{\omega}
\newcommand{\gz}{\zeta}
\newcommand{\gp}{\pi}
\newcommand{\gps}{\psi}
\newcommand{\get}{\eta}
\newcommand{\gch}{\chi}
\newcommand{\gG}{\Gamma}
\newcommand{\gD}{\Delta}

\newcommand{\gL}{\Lambda}
\newcommand{\gS}{\Sigma}
\newcommand{\gTh}{\Theta}
\newcommand{\gO}{\Omega}

\newcommand{\gPs}{\Psi}

\newcommand{\cA}{{\cal A}}

\newcommand{\cD}{{\cal D}}

\newcommand{\ui}{{\underline i}}

\newcommand{\tA}{{\tilde A}}

\newcommand{\tF}{{\tilde F}}

\newcommand{\tW}{{\tilde W}}

\newcommand{\Tr}{\mbox{Tr}}
\newcommand{\tr}{\text{tr}}

\newcommand{\Id}{\text{\small 1}\hspace{-3.5pt}\text{1}}
\newcommand{\slashed}{\hspace{-1.1ex}/}
\newcommand{\Slashed}{\hspace{-1.4ex}/\hspace{.2ex}}

\newcommand{\ra}{\rightarrow}

\newcommand{\der}{\partial}

\newcommand{\inv}{^{-1}}
\newcommand{\dsp}{\displaystyle}

\newcommand{\ubar}[1]{{\underline{#1}}}

\newcommand{\labl}[1]{\label{#1}}
\newcommand{\half}{\frac 12 }

\newcommand{\beq}{\begin{equation}}
\newcommand{\eeq}{\end{equation}}
\newcommand{\barr}{\begin{array}}
\newcommand{\earr}{\end{array}}
\newcommand{\equ}[1]{\begin{gather} #1 \end{gather}}
\newcommand{\equa}[1]{\begin{align} #1 \end{align}}

\newcommand{\tabu}[2]{\begin{tabular}{#1} #2 \end{tabular}}
\newcommand{\arry}[2]{\begin{array}{#1} #2 \end{array}}

\newcommand{\pmtrx}[1]{\begin{pmatrix} #1 \end{pmatrix}}

\newcommand{\non}{\nonumber}
\newcounter{oldcounter}

\newcommand{\fZ}{\mathfrak{ Z}}
\newcommand{\bq}{{\bar q}}

\newcommand{\bz}{{\bar z}}

\newcommand{\bR}{{\bar R}}

\newcommand{\bgth}{{\bar\theta}}

\newcommand{\bgps}{{\bar\psi}}

\newcommand{\bgPs}{{\bar\Psi}}

\newcommand{\tgg}{{\tilde \gamma}}

\newcommand{\tgs}{{\tilde \sigma}}

\newcommand{\tgG}{{\tilde \Gamma}}

\newcommand{\tgO}{{\tilde \Omega}}

\newcommand{\Intr}{\mathbb{Z}}
\newcommand{\Cplx}{\mathbb{C}}

\newcommand{\ba}[2]{\[\begin{array}{#2}\label{#1}}
\newcommand{\ea}{\end{array}\]}
\newcommand{\be}{\begin{equation}}
\newcommand{\ee}{\end{equation}}
\newcommand{\bea}{\begin{eqnarray}}
\newcommand{\eea}{\end{eqnarray}}

\newcommand{\E}[1]{\mathrm{E_{#1}}}
\newcommand{\U}[1]{\mathrm{U(#1)}}
\newcommand{\SU}[1]{\mathrm{SU(#1)}}
\newcommand{\SO}[1]{\mathrm{SO(#1)}}

\newcommand{\rep}[1]{\mathbf{#1}}
\newcommand{\crep}[1]{\overline{\rep{#1}}}

\begin{document}

%
%
\begin{flushright}
hep-th/0208146
\end{flushright}
\vskip 2 cm
\begin{center}
{\Large {\bf Localized anomalies in heterotic orbifolds} 
}
\\[0pt]

\bigskip
\bigskip {\large
{\bf F.\ Gmeiner$^{a,}$\footnote{
{{ {\ {\ {\ E-mail: gmeiner@th.physik.uni-bonn.de}}}}}}}, 
{\bf S.\ Groot Nibbelink$^{a,}$\footnote{
{{ {\ {\ {\ E-mail: nibblink@th.physik.uni-bonn.de}}}}}}}, 
{\bf H.P.\ Nilles$^{a,}$\footnote{
{{ {\ {\ {\ E-mail: nilles@th.physik.uni-bonn.de}}}}}}},
{\bf M.\ Olechowski$^{a,b,}$\footnote{
{{ {\ {\ {\ E-mail: olech@th.physik.uni-bonn.de}}}}}}}
{\bf M.\ Walter$^{a,}$\footnote{
{{ {\ {\ {\ E-mail: walter@th.physik.uni-bonn.de}}}}}}}
\bigskip }\\[0pt]
\vspace{0.23cm}
${}^a$ {\it  Physikalisches Institut der Universit\"at Bonn,} \\
{\it Nussallee 12, 53115 Bonn, Germany.}\\
\vspace{0.23cm}
${}^b${\it  Institute of Theoretical Physics, Warsaw University,} \\
{\it Ho\.za 69, 00--681 Warsaw, Poland.}\\
\bigskip
\vspace{3.4cm} Abstract\\[2ex]
\end{center}
Recently spatially localized anomalies have been considered in higher
dimensional field theories. The question of the quantum consistency
and stability of these theories needs further discussion. Here we
would like to investigate what string theory might teach us about
theories with localized anomalies. We consider the $\Intr_3$ orbifold 
of the heterotic $\E{8}\times \E{8}'$ theory, and compute the anomaly
of the gaugino in the presence of Wilson lines. We find an anomaly
localized at the fixed points, which depends crucially on the
local untwisted spectra at those points. We show that non--Abelian
anomalies cancel locally at the fixed points  for all $\Intr_3$ models
with or without additional Wilson lines.  
At various fixed points different anomalous $\U{1}$s may be present, 
but at most one at a given fixed point. 
It is in general not possible to construct one generator which is
the sole source of the anomalous $\U{1}$s at the various fixed points.

\newpage

%
%
\section{Introduction}

Recently, there have been various calculations of the structure of 
anomalies in five dimensional field theories compactified on orbifolds,
where it was shown that the anomalies are localized at the orbifold
fixed points. The first such calculation was presented in ref.\ 
\cite{Arkani-Hamed:2001is} for $S^1/\Intr_2$, and subsequent works on
the orbifold $S^1/\Intr_2 \!\times \! \Intr_2'$ can be found 
in refs.\ \cite{Scrucca:2001eb,Pilo:2002hu,Barbieri:2002ic} using
Kaluza--Klein expansions of the five dimensional fermions and gauge
fields. In ref.\ \cite{GrootNibbelink:2002qp} topological arguments 
taken from \cite{Horava:1996qa,Horava:1996ma} were used to 
obtain the same results for Abelian and non--Abelian gauge theories. 
It turned out that if there is no anomaly for the zero modes of the
theory, then it is possible to choose a regularization scheme or 
introduce appropriate Chern--Simons terms so as to cancel the local
gauge anomaly. A localized $\U{1}$ gauge anomaly (or more precisely, a
mixed $\U{1}$ gravitational anomaly), is associated to quadratically
divergent Fayet--Iliopoulos terms \cite{Ghilencea:2001bw} localized 
at the fixed points
\cite{Scrucca:2001eb,Barbieri:2002ic,GrootNibbelink:2002qp,GrootNibbelink:2002wv}.
In \cite{GrootNibbelink:2002qp,GrootNibbelink:2002wv} 
it was shown that those localized Fayet--Iliopoulos terms give
rise to an instability, which leads to localization of charged bulk
fields to one of the fixed points of the orbifold.

This investigation of localized anomalies and their physical consequences
on five dimensional orbifolds in field theory is very interesting, but
it leads to the question whether such theories are fully consistent
quantum field theories and free of instabilities. This becomes even
more pronounced when we go to higher dimensions; some investigations
in that direction have been performed in  
\cite{Hebecker:2001jb,Asaka:2002nd}. 
One strategy of answering these questions would be to consider higher
dimensional string theories, that are believed to represent consistent
quantum theories including gravity. We could then learn what string
theory teaches us about localized anomalies. Local anomalies have been
explored before in the context of open string theory
\cite{Scrucca:2002is, Antoniadis:2002cs}, (heterotic) M--theory orbifolds 
\cite{Faux:1999hm,Faux:2000dv,Kaplunovsky:1999ia,Gorbatov:2001pw}, 
heterotic M--theory in five dimensions \cite{Lukas:1999nh}.

In order to have a framework to perform
explicit calculations of anomalies, we consider the field
theory limit (i.e.\ the limit of the string tension $\ga'\ra 0$) 
of the $\E{8}\times \E{8}'$ heterotic string compactified on an
orbifold \cite{dixon_85,Dixon:1986jc}. 
For simplicity we take this orbifold to be $T^6/\Intr_3$. 
It is well--known, that the string spectrum then contains both
so--called untwisted (bulk) and twisted (brane) states. 
The field theory limit of the
untwisted sector gives rise to ten dimensional $N=1$ supergravity
coupled to $\E{8}\times \E{8}'$ super Yang--Mills gauge theory on this
orbifold. The twisted states are localized at the fixed points of the
orbifold only. Modular invariance determines their spectrum
precisely; even when Wilson lines are present
\cite{ibanez_87,ibanez_88,Font:1988tp}.  
(This is contrary to orbifold field theories, where fixed point 
states are only required to be compatible with the (super) symmetries
present at the fixed points.)  
In addition, we will relate our localized anomaly results to the
well--known situation of four dimensional zero modes. 
The zero mode limit (i.e.\ the radii of the orbifold $R_i\ra 0$) 
leads to an effective $N=1$ supergravity coupled to gauge
and chiral multiplets in four dimensions. 
In figure \ref{fig:HetOverview} we give a schematic overview of the 
various states and limits of heterotic string theory discussed above.  

In a recent paper \cite{Gmeiner:2002ab} the structure of localized
anomalies in heterotic string models was investigated, raising the
question how localized non--Abelian anomalies can be canceled, if
the chiral twisted and untwisted zero mode states are not
anomaly free locally. 
This observation partly triggered the work reported in this
paper. Here, we give a simpler example of this situation, than in
\cite{Gmeiner:2002ab}, though the essential features are the 
same: Consider a $\Intr_3$ orbifold model with gauge shift and single
Wilson line that are equal:  
$v = a_1 = ( \mbox{-}2, ~1^2,  ~0^5 ~~|~~ 0^8)$. 
The main properties of this model are: the unbroken gauge group 
is  $\E{6} \!\times\! \SU{3} \times \E{8}'$, and there is no
untwisted matter. The twisted states at the $27$ fixed points fall
into three sets that are distinguished by the Wilson line on the first
two torus. Classifying the twisted states according to the fixed
points, denoted here by $0, 1,2$, on that two torus, we have
\equ{
0: ~~ 
(\rep{1},\rep{27})(1)' + 3\cdot(\rep{3},\rep{1})(\rep{1})', 
\quad
1: ~~ 
(\rep{1},\crep{27})(1)' + 3\cdot(\crep{3},\rep{1})(\rep{1})',
\quad 
2: ~~  
3\cdot(\bar{\rep{1}},\rep{1})(\rep{1})'.
\labl{twistedstatesv=a}
}
In this example it is clear that the global four dimensional anomaly
cancels; but there seems to be an $\SU{3}$ anomaly locally at the
fixed points $0$ and $1$, as there are no chiral untwisted states that
can compensate the chiral twisted states at the fixed points. Does
this mean that in this model we observe localized non--Abelian
anomalies? 

In this work we will show that the answer to this question is no. 
There are no localized non--Abelian gauge anomalies on the heterotic
$\Intr_3$ orbifold. The situation with Abelian gauge anomalies is more
complex and will be clarified in section \ref{sect:anomU1}.
\\[-1ex]

\noindent 
The remainder of this paper is organized as follows:
\newcounter{sctns}
\begin{list}{Section \arabic{sctns}.}
{\usecounter{sctns}
\setlength{\labelwidth}{0ex}
\setlength{\labelsep}{2ex}
\setcounter{sctns}{1}
}
\item Field theory description of the heterotic string on
$T^6/\Intr_3$.
\\[1ex]
The basic features of the field theory approximation of heterotic
$\E{8}\times \E{8}'$ string theory on the orbifold $T^6/\Intr_3$ are
recalled; in particular, the Hosotani description of Wilson lines in subsection
\ref{sect:yangmills}.  The untwisted matter surviving the orbifold
twist projection at a given fixed point are determined in subsection
\ref{sect:untwistedFixedZero}; and its relation to the four
dimensional zero mode matter is indicated. The final subsection
\ref{sect:twistedhet} discusses the additional twisted states at the
fixed points. 
\item Orbifold gauge anomalies.
\\[1ex]
The core of this work is formed by the computation of the gaugino
anomaly in the presence of Wilson lines in subsection
\ref{sect:GauginoAnomaly}.  Using Fujikawa's method in ten dimensions
we show that the anomaly becomes localized at the four dimensional
fixed points.  The formula for the localized anomalies due to twisted
states is exposed next. The reduction to the four dimensional zero
mode anomaly concludes this section.   
\item Localized heterotic anomalies.
\\[1ex] 
In this section we combine the results of the previous sections to
investigate the structure of anomaly cancelation in heterotic string
theory.  To facilitate this discussion we introduce the concept of 
fixed point equivalent models. 
\\
Using this concept it is straightforward to infer that non--Abelian
gauge anomalies always cancel locally, even in the presence of Wilson
lines. In subsection \ref{sect:v=a} we return to the example 
\eqref{twistedstatesv=a} of this introduction and explain the
resolution to the apparent paradox. 
\\
The localized anomalous $\U{1}$s have a richer structure as is
explained in subsection \ref{sect:anomU1}. Some intriguing aspects of
the localized anomalous $\U{1}$s are illustrated in examples
in section \ref{sect:ExampleU1}.  
\item Conclusion and outlook.  
%
\item[Appendices.] ~~~~
\\[1ex]
Various more technical issues, the decomposition of the ten
dimensional Clifford algebra, the torus gaugino wave functions, and 
the action of Weyl reflections on shift vectors,  are collected. 
\end{list}

\begin{figure}
\begin{center}
\raisebox{0mm}{
\scalebox{0.8}{\mbox{\input{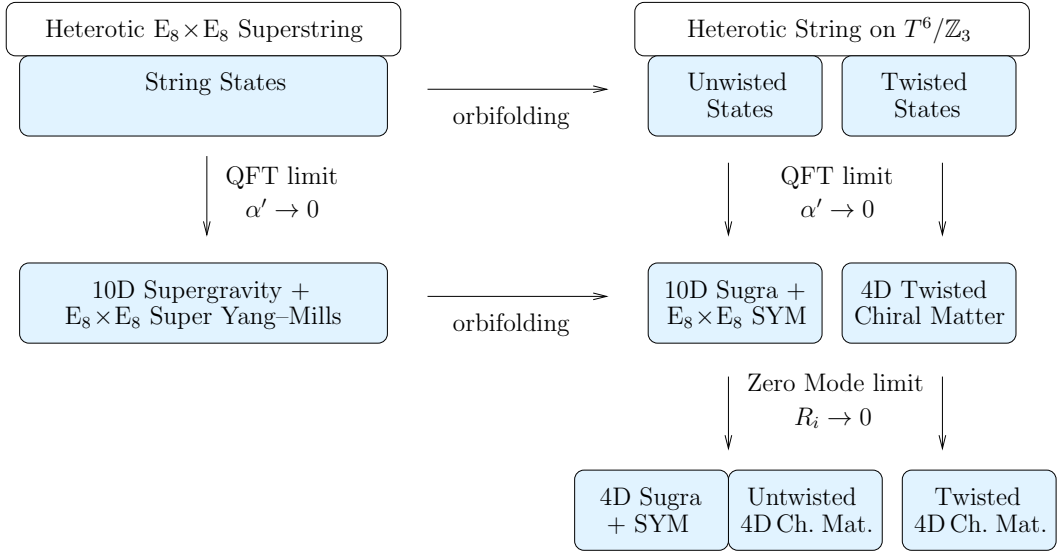}}}}
\end{center} 
\caption{The figure shows that orbifolding and taking the low energy
field theory limit ($\ga' \ra 0$) of the heterotic string theory do
not commute. If the limit is taken first, then ten dimensional
supergravity coupled to super Yang--Mills is obtained. This theory
does not contain the additional twisted string states that arise by
putting the string on the orbifold. By taking all radii 
$R_i \ra 0$ the theory becomes an effective $N=1$ supersymmetric field
theory in four dimensions.} 
\labl{fig:HetOverview}
\end{figure}

%
%
\section{Field theory description of the heterotic string on $\boldsymbol{T^6/\Intr_3}$}
\labl{sect:heterotic}

In this section we set the stage for our investigation of the
structure of localized anomalies in heterotic string models
compactified on orbifolds. 
String orbifold models without Wilson lines were studied for the first
time in \cite{dixon_85,Dixon:1986jc}, some useful tables can be found
in \cite{Katsuki:1989kd,Katsuki:1990bf}. 
The coordinates of the four dimensional Minkowski space are denoted by
$x$.  The six internal $T^6$ torus coordinates are combined to form
three complex planes described by the complex numbers 
$z = (z^i) = (z^1,z^2,z^3)$ (and with $\bz = (\bz^\ui)$ we denote
their complex conjugates). These complex coordinates transform as a
$\rep{3}_H$ with respect to the holonomy group $\SU{3}_H$, 
embedded in the $D=(1,9)$ Lorentz group. (The subscript $_H$ is used
to distinguish the holonomy $\SU{3}_H$ representations, from the
representations that are related to the $\E{8} \times \E{8}'$ gauge
group.) 

The two dimensional tori are defined by the
relations $z^i \sim z^i + R_i$ and $z^i \sim z^i + e^{2\pi i\, \gf_i} R_i$.  
Here the phases $e^{2\pi i\, \gf_i}$ are defined in terms of 
numbers $\gf_i$ that satisfy $3 \gf_i \equiv 0$ (where the equivalence
relation $a \equiv b$ means $a$ is equal to $b$ modulo an integer.) 
If we use $\hat \imath, e^{2\pi i\, \gf_i} \hat \imath$ as the basis
vectors of these three complex tori with length $R_i$, an integral
lattice $\gG$ can be defined by 
\(
\gG = \{ m_i \, \hat \imath + n_i \, e^{2\pi i\, \gf_i} \hat \imath |
m_i, n_i \in \Intr \}.
\)
On each of these complex planes the $\Intr_3$--orbifold twist operator
$\gTh$ acts as an $\SU{3}_H$ element
\equ{
\gTh(x, z^1, z^2, z^3) = 
(x, e^{2\pi i\, \gf_1} z^1, e^{2\pi i\, \gf_2} z^2, e^{2\pi i\, \gf_3}z^3), 
\labl{OrbiTwistCoor}
}
so that $\sum_i \gf_i \equiv 0$. 
The definition of the orbifold--twist is extended to spinors in
ten dimensions by requiring that 
\equ{
-\! \gG_0 R_\gTh^\dag \gG_0 \,\gG_M\, R_\gTh 
=(\gTh\inv)_M{}^N   \gG_N,
\qquad 
R_\gTh^3 = \Id. 
}
For the properties and conventions concerning the Clifford algebra
that we are employing, see appendix \ref{SpinorDims}. An explicit
representation of $R_\gTh$ reads 
\equ{
R_\gTh = e^{-2\gp\, \gf^j\, \half \gS_j},
\qquad 
\gS_j = \frac 12 [ \gG_{2j+2}, \gG_{2j+3}] = 
\begin{cases}
i \Id \otimes \Id \otimes \gs_3 & j = 1, \\
i \Id \otimes \gs_3 \otimes \Id & j = 2, \\
i \gs_3 \otimes \Id \otimes \Id & j = 3. 
\end{cases}
\labl{SpinTwist}
}
Of course, the central importance of the orbifolding lies in the fact
that it breaks $3/4$ of supersymmetries in the effective four
dimensional theory. The requirement for the existence of a spinorial 
supersymmetry parameter $\ge(x)$ which is constant over the orbifold
and satisfies  
\(
R_\gTh \ge(x) = \ge(x), 
\)
can be solved in terms of one four dimensional Majorana spinor. 
Using the notation $\get^\ga = \get^{\ga_1\ga_2\ga_3}$ for an six
dimensional spinor, with two dimensional chiralities $\ga_i$, as
introduced in  appendix \ref{SpinorDims}, we find that 
\equ{
R_\gTh \get^{\ga} = e^{-2\pi i\, \frac 12 \gf_i\ga_i}\get^{\ga}. 
}
The numbers $\gf_i$ can be chosen such that $\get^0 = \get^{+++}$ is
invariant, while  the other spinors 
$\get^a = (\get^{+--}, \get^{-+-}, \get^{--+})$ 
form a triplet $\rep{3}_H$ of $\SU{3}_H$ and transform with the same
phase.  (The charge conjugate spinors transform with opposite
phases.) A consistent choice for the numbers $\gf_i$, which we use from
now on, is such that all phases $\gth = e^{2\pi i\, \gf_i}$ are equal:
$(\gf_1, \gf_2, \gf_3) = \frac 13 (1, 1, -2)$.

In each of the three complex tori we have three fixed points,
therefore we have 27 fixed points in total which are collectively
denoted by 
\(
\fZ_s  = ( R_1 \gz_{s_1}, R_2 \gz_{s_2}, R_3 \gz_{s_3} ) 
\)
with the integers $s = (s_1, s_2, s_3)$, $s_i = 0,1,2$. Here we have
defined the fixed points $\gz_0 = 0,  \gz_1 = \frac 13(2 + \gth)$ and 
$\gz_2 = \frac 13(1 + 2 \gth)$ for a unit two dimensional $\Intr_3$
orbifold, which satisfy 
\equ{
\qquad 
 \gth \, \gz_0 = \gz_0, 
\qquad 
\gth \,\gz_1 = \gz_1 \, - 1, 
\qquad 
\gth\,  \gz_2 = \gz_2 \, - 1 - \gth.
\labl{fixedpoints}
}
The delta function on the torus $\gd(z - \gG)$, obtained from
the delta function $\gd(z)$ on $\Cplx^3$, has the property that 
\equ{
\gd((1-\gth^k)z-\gG) = \sum_s \frac 1{27}\gd( z - \fZ_s - \gG),
\labl{deltaOrbi}
}
for $k = 1, 2$. The factor $1/27$ arises because of the factor
$(1-\gth^k)$ in front of $z$ in the first delta function.

\subsection{Super Yang--Mills and supergravity on 
$\boldsymbol{T^6/\Intr_3}$}
\labl{sect:yangmills}

We now give a brief recapitulation of the field content of  the
supersymmetric low energy field theory of the heterotic $\E{8} \times
\E{8}'$ string in $D =(1,9)$ dimensional Minkowski space. The Lagrangian
for this supergravity theory coupled to super Yang--Mills gauge 
theory has been first derived in ref.\
 \cite{Chamseddine:1981ez,Bergshoeff:1982um,Chapline:1983ww}
(for a text book introduction, see \cite{gsw_2}). 
The supergravity multiplet contains the metric $G_{MN}$, the
anti-symmetric two form $B_{MN}$ and a dilaton $\gf$ as bosonic 
states, and the left--handed Majorana--Weyl Rarita--Schwinger field
$\gps_M$ and a right--handed Majorana--Weyl spinor $\gl$ as 
fermionic states. The gauge multiplet consists of gauge fields $A_M$
and left--handed Majorana--Weyl gauginos $\gch$ both in the 
adjoint of $\E{8} \times \E{8}'$. As we will be primarily concerned
with gauge anomalies, we only state the gauge transformation
of the gauge multiplet
\equ{
i {\;}^g \!A_M(x, z) = g(x, z) \Bigl( i A_M(x,z) + \der_M \Bigr) g\inv(x, z), 
\qquad 
{\;}^g \!\gch(x, z) =  g(x, z) \gch(x,z)  g\inv(x, z),
\labl{Gauge}
}
with the local $\E{8}\times \E{8}'$ gauge group element $g(x,z)$.
(For most parts of the discussion below it is irrelevant that the gauge
group of the heterotic theory is a direct product of two simple Lie
groups.)

Next we describe how the orbifold boundary condition is realized on
the gauge multiplet. (The analysis for the supergravity multiplet is
analogous.) First the gauge field is discussed, as this is more involved
than the situation with the gaugino. Let $H_I$ denote the generators
of a Cartan subalgebra of the gauge group $\E{8} \times \E{8}'$, and
denote by $E_{w}$ the other generators with weights (roots) $w$. 
Assume that these Hermitean generators are normalized such that 
\equ{
[ H_I, E_w ] = w_I E_w, 
\qquad 
H_I^\dag = H_I^{\;}, 
\qquad E_w^\dag = E^{\:}_{-w},
\qquad
e^{2\pi i\, v^I H_I}\, E_w\, e^{-2\pi i\, v^I H_I} = e^{2\pi i\, v^I w_I}
E_w \; . 
\labl{Algebra}
}
The algebra valued gauge fields can be decomposed  as 
$A_M = A_M^I H_I + A^w_M E_w$, where the sum over 
the Cartan index $I$ and the roots $w$ is understood. 
Using the collective notation for the generators $T_A$, the Killing
metric is defined as $\get_{AB} = \frac 1{2g}\tr(T_A^\dag T_B^{\;})$, 
($g$ denotes the dual Coxeter number of $\E{8}\times \E{8}'$) 
and its inverse is denoted by the indices as superscript. 
The orbifold--twist acts on the gauge fields as 
\equ{
A_M(x, \gTh z) = (\gTh\inv)_M{}^N U A_N(x, z) U\inv,
\qquad 
U = e^{2\pi i\, v^I H_I} 
\quad \text{with} \quad 
\forall w:~ 3 v^I w_I \equiv 0.
\labl{GaugeOrbiTwist}
}
The last condition assures that the matrix $U$ is a $\Intr_3$
representation. Because the gauge field can be represented as a
one--form  $A = A_M \d x^M$, it follows from \eqref{OrbiTwistCoor}
that $A_M$ transforms with the inverse twist. 

The gauge fields do not have to be strictly invariant when going round
one of the cycles of the torus $T^6$. Indeed, they only need to be
invariant up to a group transformation. Since the cycles of the torus
are non--contractable, the Hosotani mechanism \cite{Hosotani:1989bm} 
can be at work by implementing Scherk--Schwarz boundary
conditions \cite{Scherk:1979ta} using the gauge symmetry. This gives
rise to three matrices $T_i$ with the properties 
\equ{
A_M(x, z + \hat \imath) = T^{\,}_i A_M(x, z) T_i\inv,
\quad 
A_M(x, z + \gth\, \hat \imath) = T^{\,}_i A_M(x, z) T_i\inv,
\qquad 
T_i = e^{2 \pi i\, a_i^I H_I},
\labl{ScherkSchwarz}
}
with $\forall w:~ 3 a_i^I w_I \equiv 0$. For simplicity we have chosen
the orbifold and the Hosotani boundary conditions to commute, and
reside in the Cartan of $\E{8} \times \E{8}'$. 
The consistency condition on the Wilson line coefficients $a_i^I$
follows directly because $1 + \gth + \gth^2 = 0$.

Instead of working with a gauge field $A_M$, which is periodic
on the torus $T^6$ up to a group transformation, one can also work
with a strictly periodic gauge field $\tA_M$, by performing the
following field redefinition 
\equ{
A_M(x, z) = T(z) \tA_M(x, z) T\inv(z), 
\quad 
T(z) = e^{2\pi i\, a^I(z) H_I},
\quad
a^I(z) =  
\frac {z^i\, (1 - \bgth) - \bz^\ui \, (1 - \gth)}{\gth - \bgth} 
\frac 1{R_i} a_i^I. 
\labl{WilsonMat}
}
Notice that the functions $a^I(z)$ are real: $(a^I(z))^* = a^I(z)$. 
From the shift property 
\(
a^I (z + \hat \imath) = a^I(z + \gth \hat \imath) = a^I(z) + a_i^I,
\)
for $i = 1,2,3$ it follows directly that if $\tA_M$ is periodic, 
$A_M$ satisfies \eqref{ScherkSchwarz}. However, this field
redefinition can be partly undone by a gauge transformation
\eqref{Gauge} with gauge group element $T(z)$   
\equ{
i A_M(x, z) = T(z) 
\Bigl( i \tA_M(x,z) + i B_M + \der_M \Bigl) 
T\inv(z), 
\quad 
B_i = 
2 \pi \, \der_i a^I(z) H_I  = 
 \frac {2 \pi}{R_i} \, \frac {1 - \bgth}{\gth - \bgth} \, a_i^I H_I.
}
Of course, also the conjugate $B_\ui$ is non--vanishing, while 
$B_\gm = 0$. 
The conclusion of this computation is that a theory with (gauge) field
periodic up to group transformations, generated by the Cartan
subalgebra of the gauge group, is equivalent to having truely periodic
gauge fields which have constant background values $B_i, B_\ui$ 
in the Cartan subalgebra directions. As we have seen above these
so--called Wilson--lines  are quantized because of consistency with
the orbifold boundary condition. 

Next we turn to the gauginos on $T^6/\Intr_3$. Because of gauge
invariance, the gauginos satisfy the fermionic
equivalent of the boundary conditions \eqref{GaugeOrbiTwist} 
and \eqref{ScherkSchwarz}
\equ{
\gch(x, \gTh z) = U \, R_\gTh\,  \gch(x, z) U\inv,
\qquad 
\gch(x, z + \hat \imath) = T_i \, \gch(x, z) T_i\inv,
\labl{GauginoBoundCond}
} 
with the spinor twist--rotation $R_\gTh$ given in \eqref{SpinTwist}.

\subsection{Untwisted matter at the fixed points and their zero modes}
\labl{sect:untwistedFixedZero}

In this subsection we identify all untwisted states that exist at the
fixed points. In general this set of states is larger than the set of zero
mode untwisted states that we will discuss at the end of this
subsection. Again we only focus on the gauge multiplet
here. At the fixed point $\fZ_{s}$ the surviving states have
to satisfy the condition 
\equ{
(\gTh\inv)_M{}^N U A_N(x, \fZ_{s}) U\inv =
A_M(x, \gth \fZ_{s}) = 
(T_1^{s_1} T_2^{s_2} T_3^{s_3})\inv  
A_M(x, \fZ_{s}) (T_1^{s_1} T_2^{s_2} T_3^{s_3}). 
\labl{FixedGaugeCond}
}
In the first equality we have used the action of the orbifold--twist
given in \eqref{GaugeOrbiTwist}. The second equality follows 
because the functions $a^I(z)$ introduced in \eqref{WilsonMat} have
the property that for any integer $k$ 
 \equ{
a^I(\fZ_s) - a^I(\gth^k \fZ_s) \equiv k s_i a_i^I,
\labl{WilsonLines}
}
using the definition of the fixed points \eqref{fixedpoints}. 
This equation can also be stated as the projection for both the gauge
fields and the gauginos  
\equ{
\arry{r}{ 
(\gTh\inv)_M{}^N 
U_{s} A_N(x, \fZ_{s}) U_{s}\inv =  A_M(x, \fZ_{s}), 
\\[2mm]
U_{s} \, R_\gTh \gch (x, \fZ_{s}) U_{s}\inv = \gch(x, \fZ_{s}), 
~~\,
} 
\qquad 
\arry{l}{
U_{s} = T_1^{s_1} T_2^{s_2} T_3^{s_3} U = 
e^{\dsp 2 \pi i Q_s}, 
\\[2mm]
Q_s = v_s^I H_I, 
\quad 
v_s^I = v^I + s_i a_i^I.
}
\labl{FixedPointCondition}
}
With the algebra \eqref{Algebra} this condition can be worked out for 
the various components, we get  
\equ{
\pmtrx{ \dsp 
A_\gm^I & A_\gm^w 
\\[2mm]
A_\ui^I & A_\ui^w 
} \!\!(x, \fZ_{s}) = 
A_M(x, \fZ_{s}) = 
\pmtrx{ \dsp 
A_\gm^I & e^{\dsp 2\pi i \, v_s^I\, w_I} A_\gm^w 
\\[2mm]
e^{{\dsp 2\pi i\,} \frac 13} A_\ui^I & 
e^{{\dsp 2\pi i(} \frac 13 +  {\dsp v_s^I\,  w_I)}} A_\ui^w 
} \!\!(x, \fZ_{s}).
\labl{GaugeOrbiTwistComp}
}
And for the gauginos we find the fixed point projections 
\equ{
\pmtrx{ 
\gch^I_0 &\gch_0^w 
\\[2mm]
\gch^I_a & \gch_a^w 
} \!\! (x,  \fZ_{s}) 
= \gch(x, \fZ_{s}) = 
\pmtrx{ 
\gch^I_0 &   e^{\dsp 2\pi i \, v_s^I \,w_I} \gch_0^w 
\\[2mm]
e^{{\dsp 2\pi i\,} \frac 13} \gch^I_a &
e^{{\dsp 2\pi i(} \frac 13  +  {\dsp v_s^I \,w_I)}} \gch_a^w 
} \!\! (x,  \fZ_{s}),
}
where we used $v_s^I$ defined in \eqref{FixedPointCondition}. 
Because of the Majorana--Weyl condition of the gauginos in ten
dimensions, it follows that the four dimensional negative chirality
states are not independent, and therefore we did not indicate them
here. 

The Cartan subalgebra gauge fields always 
exist at the fixed points. In addition we find, that the gauge fields
and gauginos, as well as the complex scalars and chiral fermions that 
exist at the fixed point $\fZ_{s}$ are determined by the relations:
\equ{
\text{gauge}~\rep{Ad}_s: 
\left\{   
\arry{l}{ 
\!\! v_s^I \, w_I \equiv 0 
\\[2mm]
\!\! H_I 
} \right. \!\!,
\  
\text{matter}  
\left\{
\arry{l}{
\!\!
(\rep{3}_H, \rep{R}_s: 
\mbox{\small $\frac 13$} +  v_s^I \,  w_I \equiv 0) 
\\[2mm]
\!\!
(\crep{3}_H, \crep{R}_s:  
\mbox{\small $\frac 23$} +  v_s^I \, w_I \equiv 0) 
}
\right. \!\!, 
\  
\text{at fixed point} ~ \fZ_{s}.
\labl{GaugeMatterFixed}
}
Here $\rep{Ad}_s$ denotes the adjoint representation of gauge fields and
gauginos at fixed point $\fZ_s$, corresponding to both the Cartan
subalgebra and the generators $E_w$ which commute with 
$U_s$, defined in \eqref{FixedPointCondition}. The corresponding gauge
group is $G_s \subset \E{8} \times \E{8}'$. These gauge multiplets are
singlets under the holonomy group $\SU{3}_H$. 
The untwisted matter at fixed point $\fZ_s$ consists of chiral
multiplets containing complex scalars $A_\ui^w$ and the left--handed
fermions $\gch_a^w$; they form triplets of $\SU{3}_H$. 
We label this set of chiral multiplets by $\rep{R}_s$. 
(Their conjugates, like 
\(
A_i^w (x,  \fZ_{s}) =
\exp {{\dsp 2\pi i(} \frac 23 +  {\dsp v_s^I\, w_I)}} 
A_i^w (x,  \fZ_{s}),  
\) 
 are labeled by $\bR_s$, but they are not independent.)  
Since the union of the set $\rep{Ad}_s$, $\rep{R}_s$ and $\crep{R}_s$ 
describes the full $\E{8}\times \E{8}'$ algebra, and the group $G_s$
has maximal rank (as it contains the full Cartan of 
$\E{8}\times\E{8}'$), it follows that each $\rep{R}_s$ labels a
representation of the corresponding group $G_s$.  

If the condition for the surviving gauge group at the fixed point
$\fZ_s$ had been $v_s^I\, w_I = 0$, the group $G_s$ would always
have contained two $\U{1}$ factor generators 
\equ{
q_s = \left. Q_s \right|_{\E{8}} =  v_s^I \left. H_I \right|_{\E{8}}, 
\qquad 
q_s' = \left. Q_s \right|_{\E{8}'} =  v_s^I \left. H_I \right|_{\E{8}'}, 
\labl{U1factorGens}
}
because the heterotic gauge group is a direct product of two
$\E{8}$s. With the notation $|_{\E{8}}$ and $|_{\E{8}'}$ we indicate
that this expression is projected onto the first and second $\E{8}$ factor,
respectively. The surviving generators $\rep{Ad}_s$ have to commute
with $U_s$, rather than $Q_s$ (they only have to satisfy 
$v_s^I\, w_I \equiv 0$, instead of $v_s^I\, w_I = 0$).  Therefore, 
it may happen that the generators $q_s$ and $q_s'$ become part of a
non--Abelian factor in $G_s$. 

The four dimensional zero modes in the effective four dimensional
theory are represented by constant states over the full internal
dimension.  Therefore they exist at all fixed points simultaneously, 
so that the conditions for the separate fixed points $\fZ_{s}$ have to
be superimposed: $\forall s:~ v_s^I\, w_I \equiv 0$. From this we find 
the well-known condition for the zero mode untwisted string sector:  
\equ{
\text{gauge}~\rep{Ad}: 
\left\{   
\arry{l}{ 
\!\! v^I \, w_I \equiv a_i^I\, w_I \equiv 0 
\\[2mm]
\!\! H_I 
} \right. \!\!,
\ 
\text{matter}  
\left\{
\arry{l}{
\!\!
(\rep{3}_H, \rep{R}: 
\mbox{\small $\frac 13$} +  v^I \,  w_I \equiv a_i^I\, w_I \equiv 0) 
\\[2mm]
\!\!
(\crep{3}_H, \crep{R}:  
\mbox{\small $\frac 23$} +  v^I \, w_I \equiv a_i^I\, w_I \equiv 0) 
}
\right. \!\!, 
\ 
\text{for 4D zero modes,}
\labl{GaugeMatterZeroModes}
}
for all $i = 1, 2, 3$. This implies in particular that 
$G = \cap_s G_s$.

\subsection{Twisted heterotic string states}
\label{sect:twistedhet}

As discussed in the introduction and indicated in figure
\ref{fig:HetOverview} the existence of additional twisted string  
states cannot be inferred from ten dimensional supergravity coupled
to super Yang-Mills theory. Therefore, we briefly review in this
subsection how the zero modes of the twisted string states can be 
derived. Our presentation is based on \cite{ibanez_87}, where the
inclusion of Wilson lines in orbifold models was studied for the first
time. (See also \cite{Casas:1989wu} for the $\Intr_3$ orbifold
specifically.)  

First, we would like to stress, that for modular invariance of 
string theory we have to fulfill another consistency condition for every
fixed point $s$, namely
\begin{equation}
\labl{eq:modinv}
\mbox{$\frac 32$}(v_s^I)^2 \equiv 0.
\end{equation}
As we are only interested in the
QFT limit $\alpha' \rightarrow 0$ it suffices to examine the massless
twisted states. However, they are necessarily localized at the fixed
points, which follows from the oscillator expansion and
monodromies of the corresponding coordinate fields. 
This implies, that they are unaffected by the zero mode
limit $R_i \rightarrow 0$, since they do not depend on the size of the
tori. Therefore, we can take the massless twisted chiral
matter, as computed from string theory, and include it in the field
theory analysis of the gauge anomalies presented in the next section. 

The twisted
massless states are built up by taking the tensor product of left--
and right--moving massless states. For the $\mathbb{Z}_3$
orbifold the situation is particularly simple, since there are only two
right--moving twisted massless states, the non--degenerate vacua 
from the $\widetilde{NS}$ and the $\widetilde{R}$ sector, which we
denote by 
$| \, 0 \,\rangle_{\widetilde{NS}, \textrm{tw}}$ and
$|\,0\,\rangle_{\widetilde{R}, \textrm{tw}}$, respectively. For the
left--movers the situation is more involved. At a given fixed point
$\fZ_s$ the massless twisted states have to fulfill the condition
\equ{
\label{eq:condleft}
\mbox{\small $\frac{1}{2}$} 
\Bigl( w^I + v_s^I \Bigr)^2 
+ N - \mbox{\small $\frac{2}{3}$} = 0.
}
Here $w^I$ are elements of the $\E{8} \times \E{8}'$ root lattice and
$N$ counts (fractional) string excitations. There are two
possibilities to obtain massless states: either $N=0$ or $N=1/3$. For
the latter option, the right--moving vacua have to be exited by the
creation operators  $\alpha_{-1/3}^{i}$, and therefore these states form
a $\rep{3}_H$ triplet of $\SU{3}_H$. The obtained massless
twisted string states  
\equ{
N = 0: ~ \Bigl( 
| \rep{S}_s, 0 \rangle_{\widetilde{NS}, \textrm{tw}} , ~ 
| \rep{S}_s,  0 \rangle_{\widetilde{R}, \textrm{tw}}
\Bigr), 
\qquad 
N = \mbox{\small $\frac{1}{3}$}: ~ \Bigl(
\alpha_{-1/3}^{i} | \rep{T}_s, 0 \rangle_{\widetilde{NS},\textrm{tw}}
, ~ 
\alpha_{-1/3}^{i} | \rep{T}_s, 0 \rangle_{\widetilde{R}, \textrm{tw}}
\Bigr),
}
are labeled by the sets $\rep{S}_s$ and $\rep{T}_s$ defined by 
\equ{
\text{twisted matter} 
\left\{
\arry{l}{
\!\!
(\rep{1}_H, \rep{S}_s: 
(w^I + v_s^I)^2 =  \mbox{\small $\frac 43$} ) 
\\[2mm]
\!\!
(\rep{\bar{3}}_H, \rep{T}_s: 
(w^I + v_s^I)^2 =  \mbox{\small $\frac 23$} ) 
}
\right. \!\!.
\labl{twistedmatter} 
}
The complex conjugated states, coming from the inversely twisted
sector, are not independent and therefore not considered here.


%
%
\section{Orbifold gauge anomalies}
\labl{sect:OrbiAnomaly}

Before we embark on the calculation of the gauge anomaly of the
gaugino on the orbifold in the presence of Wilson lines, we would like
to make a few comments on some recent calculations of gauge  
anomalies in five dimensional orbifold theories (see 
\cite{Arkani-Hamed:2001is} for $S^1/\Intr_2$, and 
\cite{Scrucca:2001eb,Pilo:2002hu,Barbieri:2002ic} for 
$S^1/\Intr_2 \!\times \! \Intr_2'$), which served as an
inspiration and partial guideline for our computation. 

Many of these calculations of the
anomaly on those orbifolds have been performed using the gauge choice, 
that sets the fifth component of the gauge field to zero: $A_5 =0$. This
is a point of potential concern \cite{Scrucca:2001eb}, since one is
investigating violation of gauge invariance in a specific gauge. 
In two or more internal dimensions (like the six dimensional case
studied in this work)  the gauge choice putting all internal gauge
potentials to zero is inconsistent in general. It is only possible
when the corresponding internal field strength vanishes: 
$A_i = \der_i \gL ~\Ra ~ F_{ij} = 0$. 
In appendix B of \cite{Barbieri:2002ic} an argument is presented, that 
the result for the anomaly is independent of this gauge choice for the
five dimensional orbifold models. Using similar ideas as presented
there,   we show below, that also in our setting, with six internal
dimensions, no assumptions need to be made concerning the gauge field
in the extra dimensions. 

Any appearance of a 10D anomaly is canceled by the standard
Green--Schwarz mechanism, and will therefore be disregarded in most
parts of this paper.

A final general comment concerning the method we use in the
computation below: Instead of explicitly constructing and using the
orbifold wave functions, we introduce an orbifold projector,
which projects on orbifold consistent states. (This method has also
been used in a string theory calculation of the Fayet--Iliopoulos term
in type I orbifolds \cite{Poppitz:1998dj}.) This allows us to use
mode expansions on the torus, rather than on the orbifold, which make
computations considerably easier. This method may also be applied to
other orbifolds, for example the ones mentioned above. 

A related calculation of localized anomalies in six 
dimensions has recently appeared in \cite{Asaka:2002my}.

\subsection{Gaugino anomalies}
\labl{sect:GauginoAnomaly}

We start with the calculation of the gauge anomalies of the gaugino on the
orbifold $T^6/\Intr_3$ by reviewing the standard functional methods
\cite{Nakahara:1990th} to describe their origin.
We consider here only anomalies in gauge symmetries, but
take into account the effect of the spin connection $\go(e)$ as a
function of the vielbein $e$ in the Dirac operator.  
The Dirac operator of the gaugino maps the Hilbert
space of positive chirality to that of negative chirality. By
introducing a non--interacting right--handed fermion $\gx$, 
a Dirac operator $D\Slashed$ can be obtained for the Dirac spinor
$\gPs = \gch+ \gx$, that maps the total Hilbert space to itself.  
The classical action is gauge invariant
\equ{
S[\gPs, A,e] = - \int \d^{10} x\,e\inv\, \frac 12 \, \bgPs D\Slashed \gPs 
\quad \text{with}\quad
D\Slashed = \der\slashed + (i A\Slashed + \go\Slashed) \frac {1 + \tgG}{2},
\quad
S[ {\,}^g\! \gPs, {\!}^g\! A, e] = S[ \gPs, A, e].
}
We have an anomaly if, the effective path integral $Z[A]$ obtained by
integrating out the fermions 
\equ{
Z[A] = \int \cD \bgPs \cD \gPs e^{i S[\gPs, A, e]} 
\neq Z[{\;}^g\! A] = Z[A] e^{i \cA[g]},
\qquad 
\cA[\gL] = \Tr [ P_\gTh \gL \tgG ],
\labl{AnomalyTrace}
}
is not gauge invariant: $\cA[g] \neq 0$. 
In the last equation we have restricted ourselves to infinitesimal gauge
transformations, denoted by $\gL$. 
The trace $\Tr$ here is formal, as it is taken over both spinor and gauge
representations, as well as (over)countable states, and therefore
requires regularization. 

All this is standard, except that we have introduced the orbifold
projection operator $P_\gTh$, that projects on gaugino torus states
that are consistent with the orbifold twist \cite{Poppitz:1998dj}. 
This allows us to perform
the calculation of the anomaly on the orbifold, while working
with torus modes which are easier to handle.  
In the basis of gaugino mode functions $\get^\ga_{qA}(z)$, 
defined in \eqref{TorusExp} of
appendix \ref{sect:GauginoWave}, this projector takes the form 
\equ{
P_\gTh = \frac 13 \sum_{k=0}^2 \gth^{-\gs(\ga, A)k} \gTh^k 
\quad \text{with} \quad   
\gs(\ga,A) = 3 \Bigl( -\frac 12 \gf^i \ga_i + v^I w_I(T_A) \Bigr).
\labl{OrbiProj}
}
To show this, notice that by applying this operator on a torus
gaugino, a gaugino state is obtained which satisfies the orbifold
boundary condition \eqref{GauginoBoundCond} automatically.

Before we dive into the details of the gaugino
orbifold anomaly, we recall that the Wess-Zumino consistency condition
\cite{Wess:1971yu} of the anomaly action  
\equ{
\gd_{\gL_1} \cA[\gL_2] - \gd_{\gL_2} \cA[\gL_1] 
= \cA[\gL_3],
\qquad 
\gL_3 = [\gL_1, \gL_2],
\labl{WessZuminoCons}
}
fixes the structure of anomalies in terms of invariant $\gO_{2n}$ and
covariant anomaly polynomials $\gO^1_{2n}$. The construction of these
anomaly polynomial forms can be summarized by the descent equations
\cite{Zumino:1984rz} that hold for any integer $n\geq 0$
\equ{
\gO_{2n+2} = \d \gO_{2n+1}
= \text{ch}(iF) \hat A(R),
\quad 
\gd_\gL \gO_{2n+2}  = 0,
\quad 
\gd_\gL \gO_{2n+1} = \d \gO^1_{2n}(\gL).
\labl{DescentEq}
}
Here $\text{ch}$ is the Chern character and $\hat A$ is the roof 
genus of the curvature tensor $R$. Their expressions can be found in
\cite{Zumino:1984rz,Bardeen:1984pm,Alvarez-Gaume:1984ig,Nakahara:1990th},
for example.  To facilitate our calculation
below, we introduce the notation: $\tgO_{2n}^1$, which is
defined such that $\tr_\gr \tgO_{2n}^1 = \gO_{2n\, \gr}^1$,  where the
trace is taken over a $\gr$ representation of a group $G$.  

We turn to the evaluation of the formal trace formula
\eqref{AnomalyTrace} for the anomaly on the orbifold 
$T^6/\Intr_3$. We employ Fujikawa's method
\cite{Fujikawa:1979ay,Fujikawa:1980eg} to regularize this trace by
using the heat--kernel 
\equ{
\cA[\gL] = \Tr \Bigl[ P_\gTh \gL \tgG e^{- (D\slashed/M)^2}\Bigr] 
= \frac 13
\int_{M^4 \!\times\! T^6} \!\! \d^4 x \d^6 z\,  
\tr\Bigl[ P_\gTh \gL \tgG e^{-(D\slashed/M)^2} 
\gd(x - x') \gd(z - z' - \gG)\Bigr],
}
where the limits $x' \ra x$, $z' \ra z$ and $M \ra \infty$ are to be taken once the
anomaly expression is well--defined. The factor $1/3$ takes into
account that the volume of the orbifold is $1/3$ of that of the
torus. (In the remainder of the calculation we drop the notation 
$M^4 \!\times\! T^6$.) The trace $\tr$ is taken over all spinor and
gauge indices. The Minkowski space delta function can be expanded into
plane waves $e^{ipx}$,  while the torus delta function can be replaced by the
completeness relation \eqref{DecompIdentity} of appendix
\ref{sect:GauginoWave}. Using the action of the orbifold twist
operator $\gTh$ in the projector $P_\gTh$, one obtains 
\equa{
\cA[\gL] = 
\frac 1{2}\frac1{3^2} \int \d^4 x \d^6 z\,  
\int \frac {\d^4 p}{(2\pi)^4} &  \sum_{k, \ga, q, A, A'} \, 
\gth^{-\gs(\ga, A)k} e^{-i px'} \, 
\non \\[2mm] & 
\get^{A'A}\,  \tr_G  \Bigl[
\get^{\ga\; \dag}_{qA'} (\gth^{-k} z') \, 
\tr_4 \bigl( \gL \tgg \tgs  e^{-(D\slashed/M)^2 } \bigr) \, 
\get^\ga_{qA}(z) \Bigr] e^{ipx}. 
}
Here $\tr_G$ denotes the trace over the gauge group. The operator
$\gTh$ acts to the right, and therefore, in particular, on the Dirac
operator. To avoid having to indicate the phase $\gth^k$ all over the
place, we have implicitly performed the coordinate transformation 
$z \ra \gth^{-k} z$.  

We first investigate the case $k = 0$. The Dirac operator squared reads 
\(
D\Slashed^2 = D^2 + i \frac 14 F^{MN} [\gG_M, \gG_N]. 
\)
The eigenvalues of the operator $D^2$ are
$-(p^2 + 4 (2\gp)^2 |q_i + b_{iA}|^2/R_i^2)$ using the mode functions
$\get^\ga_{qA}(z) e^{ipx}$, 
where $b_{iA} = (1- \bgth) a_i^I w_I(T_A)/(\gth - \bgth)$.
All internal spinor components are treated equally, 
therefore $\half \sum \get^\ga {\get^\ga}^\dag = \Id$. 
 The $k = 0$ contribution can be written as  
\equ{
\cA_{k=0}(x,z) = \frac 1{3^2} \int \frac {\d^4 p}{(2\gp)^4} 
\sum_A
\tr_{10} \Bigl[ 
\tgG e^{ - \frac 14 F^{MN} [\gG_M,\gG_N]/ M^2} 
\Bigr]_{A}^{~~ A}
\non \\  
\frac 1{R_1^2R_2^2R_3^2} 
\Bigl( \frac 2{|\bgth - \gth|} \Bigr)^3 
\sum_{q}
e^{-(p^2 + 4 (2\gp)^2 |q_i + b_{iA}|^2/R_i^2)/M^2 }, 
}
where we have used the inverse Killing metric $\get^{AA'}$   
and the normalization of the wave functions. 
The trace $\tr_{10}$ over full ten dimensional spinors 
is non--vanishing if the exponential inside is expanded to fifth order. 
The resulting factor $1/M^{10}$ is partly canceled by a 
rescaling $p \ra M p$. In the limit  $M \ra \infty$ the remaining
factor $1/M^6$ is used to replace the sum by an integral: 
\equ{
\frac 1{M^6} \sum_{q} 
F\left( \frac{|q_i + b_{i\, A}|^2}{R_i^2 M^2} \right) 
\ra 
\int \d^3 n \d^3 m\,  
F\left( \frac{| \bgth n^i - m^i |^2}{R_i^2 |\bgth - \gth|^2} \right).   
}
By the change of variables 
\(
P_R^i + i P_I^i = {4\gp} ( \bgth n^i - m^i ) / (R_i |\bgth - \gth|), 
\)
this gives a six dimensional Gaussian integral. 
Hence the ($k=0$) anomaly finally reads 
\equ{
\cA_{k = 0}[\gL] = \frac 13 \int_{M^4 \!\times \! T^6/\Intr_3} 
\d^4 x \d^6 z \, \gO^1_{10|\E{8}^2}(\gL; A, F, R),  
}
where $\gO^1_{10}$ is defined by the descent equations. 
As the cancelation of this anomaly involves the well--know
Green--Schwarz mechanism in ten dimensions, we do not 
discuss this contribution in the subsequent sections.

Next we turn to the case $k \neq 0$, where the chiralities are not
treated equally. Since we work with a basis that is diagonal with
respect to the 
two--tori chiralities, the six dimensional chirality $\tgs$ is equal to 
$(-)^\ga$, see \eqref{TorusChirality}. Moreover, the Majorana condition 
allows us to restrict the sum of the internal chiralities to
$(-)^\ga = +$, removing the double counting and hence the factor
$\frac 12$. This expression can be evaluated further by
substituting the expressions for the mode functions and their
conjugates \eqref{TorusExp}, using the scalar completeness relation
\eqref{TorusComplete} to remove the sum over the torus 
momentum $q$, and that spinors $\get^\ga$ are normalized to unity: 
$(\get^\ga)^\dag \get^\ga = 1$ (no sum over $\ga$ here). 
We evaluate the trace over the gauge group, which means taking out the 
$AA'$--component of the algebra expression of the $\tgO_4^1$. 
Using the inverse Killing metric $\get^{AA'}$ we can write this as 
\equ{
\cA[\gL] =   \frac 1{3^2 } 
\sum_{k,\ga,A}   
\int \d^4 x \d^6 z\,  
\gth^{-\gs(\ga, A)k} \, 
e^{2\pi i\, (a_{A}(z) - a_A(\gth^{-k} z) )}
\left[
\tgO_4^1(\gL)(x, z) \right]_{A}^{~\, A}
\gd(z-\gth^{-k} z - \gG).
}
Because of \eqref{deltaOrbi} we only find 
non--vanishing contributions at the fixed points $\fZ_s$, and in
addition we get a factor of $1/27$. The Wilson line functions $a_A$ 
are then also evaluated at the fixed points $\fZ_s$, hence the property 
\eqref{WilsonLines} has to be used. This leads to a modification of the 
twist phase: 
\equ{
\gth^{-\gs(\ga,A,s)k} = 
\gth^{-\gs(\ga, A) k}e^{2\pi i\, (a_{A}(\fZ_s) - a_A(\gth^{-k} \fZ_s) )}
= \gth^{-\gs(\ga, A) k} e^{- 2\pi i\, k s_i a_i^I w_I(T_A)},
}
so that 
\(
\gs(\ga, A, s) = 
3\bigl( -\half \gf_i \ga_i + (v^I + s_i a_i^I) w_I(T_A) \bigr).
\)
By introducing the notation $\gs(\ga, \rep{r}) = \gs(\ga, A, s) $ for 
$T_A \in \rep{r}$, where $\rep{r} = \rep{Ad}_s,  \rep{R}_s, \crep{R}_s$
denote the different possible fixed point representations
\eqref{GaugeMatterFixed}, the anomaly can be rewritten as 
 \equ{
\cA[\gL] =   \frac 1{3^2 } 
\sum_{k,\ga,s}\  \sum_{\rep{r}= \rep{Ad}_s, \rep{R}_s, \crep{R}_s}      
\int \d^4 x \d^6 z \,  \gth^{-\gs(\ga, \rep{r})k } \, 
\left. 
\gO_{4\,  \rep{r}}^1(\gL)(x, z)  
\right|_{\rep{Ad}_s}
\frac 1{27} \gd(z - \fZ_s - \gG).
}
Observe that the anomaly polynomial $\gO_{4\, \rep{r}}^1$ is evaluated
at the fixed points $\fZ_s$, therefore the gauge parameter $\gL$, the
gauge field $A_\gm$ and field strength $F_{\gm\gn}$ are restricted to
those fixed points as well, and hence form adjoint representations
$\rep{Ad}_s$. 

The traces over the adjoint representations $\rep{Ad}_s$ never give a 
contribution for the anomaly, while the anomaly due
to $\crep{R}_s$ is opposite to that of $\rep{R}_s$: we denote this
relative sign as $(-)^{\rep{r}}$ defined by 
$-(-)^{\crep{R}_s} = (-)^{\rep{R}_s} = +$.
We need to be careful to take
the multiplicities correctly into account. For this purpose we 
collect the phase--factor multiplicities in the table below, and 
compute the sum of these phase--factors 
\equ{
\renewcommand{\arraystretch}{1.2}
\arry{lc}{ 
& ~~~~~~~ ~~~~~ \gs(\ga, \rep{r}) 
\\[1ex]
\raisebox{-2ex}{$\rep{r}$} &
\arry{c || c | c | c  }{
\text{multipl.} & ~0~ & ~ 1~  & ~2~  
\\[1mm]\hline\hline 
\rep{R}_s & 3 & 1 & 0 \\[1mm]\hline 
\crep{R}_s & 0 & 3 &1 
}
}
\qquad 
\arry{l}{\dsp 
\sum_{k,\ga,s,\rep{r} = \rep{R}_s, \crep{R}_s} 
(-)^{\rep{r}}\,  \gth^{- \gs(\ga, \rep{r}) k} = 
\\[1ex]
~~~~~~~ ~~~~~~ =3  + \gth^2 - 3\gth^2 - \gth  \, + \, 
3 + \gth -3 \gth - \gth^2= 9. 
}
\labl{PhaseCounting}
}
The first (second) four terms arise from $k=1$ ($k=2$). 
Hence we find the final result for the untwisted (or gaugino) anomaly
on the orbifold $T^6/\Intr_3$ 
\equ{
\cA_{un}  =  \int_{M^4 \!\times\! T^6/\Intr_3}
 \d^4 x \d^6 z \, 
\sum_s 3 \left. \gO_{4\, \rep{R}_s}^1(\gL; A, F, R)
\right|_{\rep{Ad}_s} 
\frac 1{27}  \gd(z - \fZ_s -\gG). 
\labl{GauginoAnomaly}
}
The factor $3$ is due to the fact that we use the integral on the
orbifold (instead of the torus).

Let us discuss a few issues of the interpretation of this result. 
We see that the gaugino anomaly (\ref{GauginoAnomaly})
takes the form of a four dimensional gauge anomaly localized at the
fixed points. Moreover, the states $\rep{R}_s$ contributing to
the anomaly at fixed point $\fZ_s$ are precisely those gaugino states 
that are not projected away at that fixed point, see
(\ref{GaugeMatterFixed}). This confirms the naive argument, that
only four dimensional gaugino anomalies can arise at the fixed points,
since only there the gaugino may give rise to a four dimensional chiral
spectrum. Moreover, the multiplicity $3$ in \eqref{GauginoAnomaly} 
is due to the fact that the chiral gaugino states at the fixed points
form a triplet under the holonomy group $\SU{3}_H$. In subsection 
\ref{sect:zeroModeAnom} we explain that also the factor $1/27$ had to
be expected.

\subsection{Twisted matter anomalies}

The situation of the gauge anomalies due to the twisted matter is 
much less involved then the one of the gauginos. As has been reviewed in 
subsection \ref{sect:twistedhet}, the twisted matter states are necessarily 
four dimensional and exist at the fixed points $\fZ_s$ of
the orbifold only. Therefore, the corresponding chiral fermions can only
give rise to four dimensional anomalies localized at the fixed
points. The total twisted anomaly reads 
\equ{
\cA_{tw}  =  \int_{M^4 \!\times\! T^6/\Intr_3}
 \d^4 x \d^6 z \, 
\sum_s 
\left[ 
\gO_{4\, \rep{S}_s}^1(\gL; A, F, R) \, +\, 
3 \gO_{4\, \rep{T}_s}^1(\gL; A, F, R)  
\right]_{\rep{Ad}_s}
\gd(z - \fZ_s -\gG),
\labl{TwistedAnomaly}
}
using the definitions of the representations $\rep{S}_s$ and 
$\rep{T}_s$ given in \eqref{twistedmatter}. The representations 
$\rep{T}_s$ contribute with a multiplicity $3$ to the anomaly,  
since they are triplets under the holonomy group $\SU{3}_H$.

\subsection{Zero mode anomalies}
\labl{sect:zeroModeAnom}

Let us conclude this section by comparing the results, obtained in the
previous subsections, to the well--known zero mode result for the
anomaly of untwisted states on the orbifold $T^6/\Intr_3$. 

For the untwisted anomaly \eqref{GauginoAnomaly} the four 
dimensional zero mode result can be obtained as follows. As was argued
above \eqref{GaugeMatterZeroModes}, the zero mode untwisted 
states are constant over the orbifold, therefore they exist at all fixed
points at the same time. All other gauge field states are massive from
the four dimensional effective field theory point of view, hence the
ten dimensional part $\cA_{k=0}$ of the anomaly does not contribute to
the zero mode anomaly. It follows
that the untwisted zero mode anomaly is given by 
\equ{
\cA_{un,zero} =  
\int_{M^4} \d^4 x \, 
\frac 1{27} 
\sum_s 3 \left. \gO_{4\, \rep{R}_s}^1(\gL; A, F, R)
\right|_{\rep{Ad}} 
= 
\int_{M^4} d^4 x \,  \left. 
3 \gO_{4\, \rep{R}}^1(\gL; A, F, R) \right|_{\rep{Ad}}. 
\label{UntwistedZeroAnomaly}
}
It should be stressed here, that the gauge fields that appear in this
formula are in the adjoint representation $\rep{Ad}$ defined in
\eqref{GaugeMatterZeroModes}. The expression after the second equality
sign is the standard result obtained by computing the zero mode
anomaly, taking into account the four dimensional chiral zero modes 
\eqref{GaugeMatterZeroModes} only. For orbifold models where the
spectrum at all fixed points is equal, this formula is easily checked;
all $27$ fixed points give the same trace over $\rep{R}_s = \rep{R}$. 

Reversing this reasoning, boils down to the argumentation of
ref.\ \cite{Horava:1996ma}. (However, when Wilson lines are present it
is not so easy to generalize such arguments to obtain the
correct form of the localized anomaly; for this reason we preformed
the direct calculation in subsection \ref{sect:GauginoAnomaly}.)

Since the twisted states are already four dimensional, the reduction
to the zero modes of the gauge fields in the anomaly formula 
\eqref{TwistedAnomaly} directly gives 
\equ{
\cA_{tw,zero}  =  \int_{M^4}
 \d^4 x  \, 
\sum_s 
\left[ 
\gO_{4\, \rep{S}_s}^1(\gL; A, F, R) \, +\, 
3 \gO_{4\, \rep{T}_s}^1(\gL; A, F, R)  
\right]_{\rep{Ad}}.
\labl{TwistedZeroAnomaly}
}


%
%
\section{Local anomaly cancelation} 
\label{sect:LocalAnomCanc}

In the previous section we have collected the various sources of gauge
anomalies in the field theory limit of heterotic $\E{8}\times
\E{8}'$ string theory on $T^6/\Intr_3$ in the possible
presence of multiple Wilson lines. Using the expressions for the
localized gauge anomalies (\ref{GauginoAnomaly}) and
(\ref{TwistedAnomaly}), due to the ten dimensional gaugino and four
dimensional chiral twisted states, respectively, one can now
explicitly check whether the anomalies cancel locally, or not. 

As was shown in the previous section, the local four dimensional 
anomalies for both the untwisted (\ref{GauginoAnomaly}) 
and twisted states (\ref{TwistedAnomaly}) are determined by the local
four dimensional chiral spectra, given in 
(\ref{GaugeMatterFixed}) and (\ref{twistedmatter}), respectively. 
Therefore the analysis of the cancelation of the localized fixed point
anomalies can be performed using the information of the four
dimensional chiral spectrum at the fixed points, only. 
The spectra for both untwisted and twisted states at fixed point
$\fZ_s$ are determined by the shift vector $v_s(v, a) = v + s_i a_i$, 
see (\ref{GaugeMatterFixed}) and (\ref{twistedmatter}).

\subsection{Fixed point equivalent models}
\label{sect:equmods}

Even if one uses the knowledge of the spectrum at the fixed points, 
the analysis of the localized gauge anomalies may not be very practical,
since the number of different models is large due to the
possibility of including Wilson lines. To overcome this hurdle, we first
develop the concept of fixed point equivalent models, which allows us to
systematically analyze the anomalies of a model with arbitrary
consistent choice of shift and Wilson lines. Fixed point equivalent
models are not only a useful tool to investigate localized anomalies,
but give a clear insight in the structure of orbifold models in general.

Consider two $T^6/\Intr_3$ orbifold models with shifts $v, v'$ and 
Wilson lines $a_i, a_i'$. Let $\fZ_s$ and $\fZ_{s'}$ be fixed points
of these orbifold models. As stressed in previous
sections, the complete spectra of (untwisted and twisted) states at
the fixed points are determined by the local shift vectors 
$v_{s}(v,a)$ and $v_{s'}(v', a')$, respectively, and therefore also the
gauge anomaly is determined by those shifts. We call these two models
fixed point equivalent, if their local gauge shift vectors are
equal up to Weyl reflections or lattice shifts; denoted by  
$v_{s'}(v',a') \simeq v_{s}(v,a)$. 
\footnote{ The concept of fixed point
equivalent models is not entirely new \cite{Kobayashi:1997pb}, however
there this concept was only introduced for the twisted states so as to
facilitate the analysis of possible low energy anomalous $U(1)$s.} 
(Properties of this equivalence relation $\simeq$ and Weyl reflections
are given in appendix \ref{sect:WeylReflect}.) 

\begin{figure}
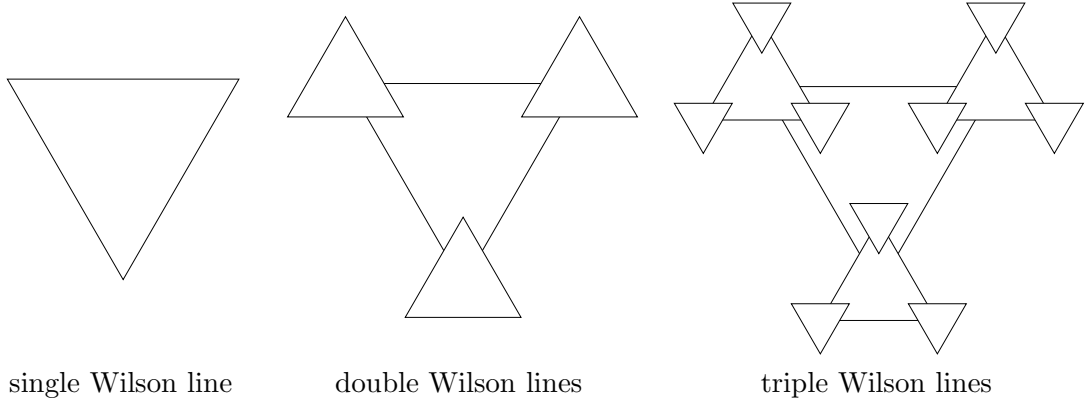

\begin{center}
\tabu{ccc}{
\raisebox{10mm}{
\scalebox{0.5}{\mbox{\input{orbif1.pstex_t}}}}
&
\raisebox{5mm}{
\scalebox{0.5}{\mbox{\input{orbif2.pstex_t}}}}
&
\scalebox{0.5}{\mbox{\input{orbif3.pstex_t}}}
\\
single Wilson line & double Wilson lines & triple Wilson lines 
}
\end{center}
\caption{For one Wilson line in an orbifold model, the fixed points in
the $T^2$ torus in which this Wilson line lies, can in principle
be distinguished from each other. In the picture on the left, this
situation is depicted: the corners of the triangle represent the $3$
sets of $9$ possibly inequivalent fixed points. Models
with two and three Wilson lines can be schematically presented 
as the middle and right diagram, respectively.  
}
\label{fig:schemaOrbi}
\end{figure}

We can also introduce the notion of equivalent fixed points $\fZ_s$
and $\fZ_{s'}$ within one model, by requiring that 
$v_{s}(v, a) \simeq v_{s'}(v, a)$. In a pure orbifold
model, of course, all $27$ fixed points are equivalent. If there is
one Wilson line, there can exist $3$ sets of $9$ fixed points that may
be inequivalent. With two Wilson lines there are in principle $9$ sets of
$3$ fixed points that may be inequivalent, and so on. With the corners
of the triangles in figure \ref{fig:schemaOrbi} the inequivalent
fixed points can be distinguished graphically. 

Using the concept of fixed point equivalent models, fixed points of
different orbifold models can thus be related to each other. In
particular, the model with gauge shift $v$ and Wilson lines $a_i$ 
at the fixed point $\fZ_s$ is equivalent to the pure orbifold theory with 
$v' = v_s(v,a)$ (and $a_i' = 0$). This naturally leads to a
classification of orbifold models in terms of pure orbifold models,
which are discussed in the next subsection. This is, of course, very 
convenient for the investigation of the localized gauge anomalies in
arbitrary orbifold models, since we can reduce the problem to a 
standard case.

\subsection{Eight pure $\boldsymbol{T^6/\Intr_3}$ orbifold models}
\labl{sect:pureOrbi}

Pure orbifold models do not have Wilson lines $v_s(v, 0) = v$,
therefore the spectra at each fixed point, in both the twisted and
untwisted sectors, are identical. Moreover, using
(\ref{GaugeMatterFixed}) and (\ref{GaugeMatterZeroModes})
it follows that local and global projections for the matter fields
are identical and local gauge groups and zero mode gauge group
are the same, i.e.\ local and global massless spectra coincide: 
$\rep{R}_s = \rep{R}$ and $G_s = G$ for all $s$. This shows that 
the zero mode anomaly (\ref{UntwistedZeroAnomaly}) is democratically
distributed over all fixed points according to the localized untwisted
anomaly (\ref{GauginoAnomaly}). Put differently, by investigating the
four dimensional zero mode anomalies, we can directly infer the
properties locally at the fixed points. This agrees with the intuitive
expectation, since without Wilson lines, the fixed points are
indistinguishable from each other. 

\begin{table}
{\small 
  \begin{center}
\renewcommand{\arraystretch}{1.25}
  \begin{tabular}{|l|l|l|ll|l}\hline
    {Model }
      & {Shift $v^I$ and }
      & {Untwisted}
      & {Twisted}
      & 
\\
& {gauge group $G = G_s$} 
& {$(\rep{3}_H,\rep{R} =  \rep{R}_s)$ }
& {$(\rep{1}_H, \rep{S}_s)$ } 
& {$(\rep{\bar{3}}_H, \rep{T}_s)$}
      \\\hline
    $\E{8}$
      & $\frac{1}{3}\!\left(~0^8 ~~~~ ~~~~~ ~~|~~ 0^8 ~~~~~ ~~~~~ \right)$
      & 
      &  
      & $(\rep{1})(\rep{1})'$
      \\
      &    $~~~~~~ ~~~~~~ ~\, \E{8}  \times   \E{8}'$
      & & & 
      \\\hline
    $\E{6}$
      & $\frac{1}{3}\!\left(\mbox{-} 2,~1^2,~ 0^5 ~~|~~ 0^8 ~~~~~ ~~~~~ \right)$
      & $(\rep{27},\crep{3})(\rep{1})'$
      & $(\rep{27},\rep{1})(\rep{1})'$
      & $ (\rep{1},\rep{3})(1)'$
      \\
       & $~~~ \E{6}\!\times\!\SU{3} \times \E{8}' ~~~~~~~ $
      & & &
      \\\hline
    $\E{6}^2$
      & $\frac{1}{3}\!\left(\mbox{-}2,~1^2,~0^5 ~~|~~ \mbox{-}2,~1^2,~0^5\right)$
      & $(\rep{27},\crep{3})(\rep{1},\rep{1}) \!+\! (\rep{1},\rep{1})(\rep{27},\crep{3})'$
      & $(\rep{1},\rep{3})(\rep{1},\rep{3})'$
      &  
      \\
      & $~~~ \E{6}\!\times\!\SU{3} \times \E{6}'\!\times\!\SU{3}'$
      &  & & 
      \\\hline
    $\E{7}$
      & $\frac{1}{3}\!\left(~0,~1^2,~0^5 ~~|~~ \mbox{-}2,~0^7 ~~~~~ \right)$
      & $(\rep{1})_{0}(\rep{64})'_{\frac{1}{2}}
 + (\rep{56})_{1}(\rep{1})'_{0}$
      & $(\rep{1})_{\frac{2}{3}}(\rep{14})'_{\mbox{-}\frac{1}{3}} $
      & $ (\rep{1})_{\frac{2}{3}}(\rep{1})'_{\frac{2}{3}}$
      \\
      &  $ ~~~~\,  \E{7}\!\times\!\U{1} \times
\SO{14}'\! \times\! \U{1}' \!\!$
      & $+\, (\rep{1})_{0}(\rep{14})'_{\mbox{-}1} + (\rep{1})_{\mbox{-}2}(\rep{1})'_{0}$
      & $ +\,  (\rep{1})_{\mbox{-}\frac{4}{3}}(\rep{1})'_{\frac{2}{3}}$ 
      & 
      \\\hline
    $\SU{9}$
     & $ \frac{1}{3}\! \left(\mbox{-}2,~1^4~,0^3 ~~|~~ \mbox{-}2,~0^7
~~~~~ \right) $
     & $(\rep{84})(\rep{1})'_{0} + (\rep{1})(\rep{64})'_{\frac{1}{2}}$
      & $(\crep{9})(\rep{1})'_{\frac{2}{3}}$
     & 
      \\
      & $~~~~~ ~~~~ \SU{9} \times \SO{14}'\!\times\!\U{1}' \!\!$
      &$+\, (\rep{1})(\rep{14})'_{\mbox{-}1}$ & &
      \\\hline
    \end{tabular}
  \end{center}
}
  \caption{The local fixed point spectra of five of the eight pure
$\Intr_3$ orbifold models are displayed. (The $\E{6}', \E{7}'$ and 
$\SU{9}'$ models are obtained from the $\E{6}, \E{7}$ and 
$\SU{9}$ models by interchanging the primed and unprimed
shift entries.) For these pure orbifold models the four dimensional
zero mode spectrum is identical to the local spectrum. }  
  \label{tab:z3models}
\end{table}
Up to the equivalence relation $\simeq$ there are exactly eight pure
$\mathbb{Z}_3$ orbifold models (cf. \cite{Dixon:1986jc},
\cite{Casas:1989wu}). Since three of those models can be obtained 
from three others by a simple interchange of $\E{8}$ and $\E{8}'$,
we only give the spectra of the five fundamental pure orbifold models
in table \ref{tab:z3models}. We have indicated the local gauge group,
the untwisted and the twisted matter at the fixed points. This table also
shows which twisted states are triplets under 
$\SU{3}_H$. Furthermore, in models with $\U{1}$ factors in the
gauge group $G$, the subscripts indicate the $\U{1}$ charges of these
matter representations.  We will use the model names $\E{8}$, $\E{6}$,
$\E{6}'$, $\E{6}^2$, $\E{7}$, $\E{7}'$, $SU(9)$ and  $SU(9)'$ to
indicate to which model the spectrum at a given fixed point in a given
model is equivalent. We use similar terminology to indicate to which
standard shift a generic shift is equivalent. In table
\ref{tab:identifyShifts} we summarized how to identify the standard
shifts; appendix \ref{sect:WeylReflect} explains how these results are
obtained. 

\begin{table}
\[
\renewcommand{\arraystretch}{1.25}
\arry{|l|ll|}{\hline
\text{number of} & v~ \text{equivalent to} & \text{corresponding}  
\\  \text{zeros in } v & \text{standard shift} & \text{gauge group}\\\hline 
8 & \frac 13\! \left(~0^8 ~~~~ ~~~~~ \right) &~~ \E{8} \\
7, 4, 1 & \frac 13\! \left( \mbox{-}2,~0^7 ~~~~~ \right) &~~ \SO{14} \\
6, 0_+ & \frac 13\! \left(~0,~1^2,~0^5\right) &~~ \E{7} \\
5, 2 & \frac 13\! \left(\mbox{-}2,~1^2,~0^5\right) &~~ \E{6} \\
3, 0_- & \frac 13\!   \left(\mbox{-}2,~1^4~,0^3 \right) &~~ \SU{9} \\\hline 
}
\]
\caption{The number of zeros (mod 1) of an $\E{8}$ shift $v$ determines to
which standard shift, given in table \ref{tab:z3models}, it is equivalent. If
the shift does not have any zeros, we distinguish whether the product
of the entries, chosen to be $\pm 1$ (using lattice shifts), is even
($0_+$) or odd ($0_-$). (For details see \mbox{appendix
  \ref{sect:WeylReflect}}.)}  
\labl{tab:identifyShifts}
\end{table}

\subsection{Non--Abelian anomalies}
\labl{sect:nonAbelAnom}

We come to our first main conclusion, using the fixed point
equivalent model analysis developed in the previous subsections. 
As has been done for the zero modes in the past, it is not difficult
to see from the spectra given in table \ref{tab:z3models}, that none of
the eight pure orbifold models has non--Abelian anomalies. 
Since any  $\Intr_3$ orbifold with Wilson lines is equivalent to one
of the pure orbifold models at a given fixed point, it follows that  
there are no non--Abelian anomalies at that fixed point. 

From this we can derive some interesting results for the anomalies in 
the effective four dimensional zero mode theory. Indeed, using
\eqref{UntwistedZeroAnomaly} we obtain 
\equ{
\Bigl[ 3\, \gO_{4\,\rep{R}}^1 + 
\sum_s \Bigl( \gO_{4\,\rep{S}_s}^1 + 3\, \gO_{4\,\rep{T}_s}^1
\Bigr) \Bigr]_{\rep{Ad}} = 
\sum_s \Bigl[ \frac 1{27} 3 \gO_{4\,\rep{R}_s}^1 
+  \gO_{4\,\rep{S}_s}^1 + 3 \gO_{4\,\rep{T}_s}^1
\Bigr]_{\rep{Ad}}. 
\labl{LocalGlobalAnomaly}
}
We see that the zero mode anomaly is the sum of the gauge anomalies at
the fixed points, with four dimensional gauge fields that do not depend
on the six internal dimensions. But since we know that there are no
non--Abelian anomalies at the fixed points, we conclude that there are
no non--Abelian gauge anomalies at the four dimensional zero mode
level. Moreover, let $Y$ be a $\U{1}$ factor generator, then it is not
anomalous at the zero mode level, if at each of the fixed points $Y$
is part of a sub--algebra that generates a non--Abelian factor in
$G_s$.  

This confirms the well--known result for the global four dimensional
gauge anomaly: 
the states subjected to (\ref{GaugeMatterZeroModes}) contribute to the
untwisted zero mode anomaly (\ref{UntwistedZeroAnomaly}). 
The zero mode twisted anomalies (\ref{TwistedZeroAnomaly}) 
are the same as the localized twisted anomalies(\ref{TwistedAnomaly}), 
and therefore determined by the spectrum \eqref{twistedmatter},
since the information concerning the localization of the twisted
states is ignored.

\subsection{Example: revisiting the $\boldsymbol{v =a_1}$ model}
\labl{sect:v=a}

In the introduction we have given one simple example of a $\Intr_3$
orbifold model where local anomaly cancelation does not seem to hold. 
However, from the analysis of the previous section, we know that for any
orbifold model the non--Abelian anomalies cancel locally. To show
exactly at which point the analysis presented in the introduction was
too naive, and to illustrate various aspects of the general investigation,
we return briefly to that example here. 

The $\Intr_3$ orbifold model discussed in the introduction has the
gauge shift equal to its single Wilson line: 
$v = a_1 = \frac 13 (\mbox{-}2,  ~1^2, ~0^5 ~~|~~ ~0^8 )$. The local
shifts $v_s$ are given by 
\equ{
\hspace{-16ex}
\arry{l}{
v_{0s_2s_3} = v ~~~~ ~~~~ = 
~\mbox{$\frac 13$} ( \mbox{-}2, ~1^2, ~0^5 ~|~ 0^8 ), 
\\[1ex] 
v_{1s_2s_3} = v+ ~\, a_1 \equiv 
\mbox{-} \mbox{$\frac 13$}(\mbox{-}2, ~1^2, ~0^5 ~|~ 0^8), 
\\[1ex]
v_{2s_2s_3} = v + 2 a_1 \equiv ~\mbox{$\frac 13$}( ~0^8 ~~~~~ ~~~~ ~|~0^8 ). 
}
\qquad 
\raisebox{-10ex}{\scalebox{0.8}{\mbox{\begin{picture}(0,0)%
\includegraphics{modelv=a.pstex}%
\end{picture}%
\setlength{\unitlength}{2763sp}%
\begingroup\makeatletter\ifx\SetFigFont\undefined%
\gdef\SetFigFont#1#2#3#4#5{%
  \reset@font\fontsize{#1}{#2pt}%
  \fontfamily{#3}\fontseries{#4}\fontshape{#5}%
  \selectfont}%
\fi\endgroup%
\begin{picture}(4046,3372)(4951,-5158)
\put(8701,-2011){\makebox(0,0)[lb]{\smash{\SetFigFont{12}{14.4}{\rmdefault}{\mddefault}{\updefault}
$v = \frac 13 (\mbox{-}2, 1^2, 0^5| 0^8)$
}}}
\put(6151,-2461){\makebox(0,0)[lb]{\smash{\SetFigFont{12}{14.4}{\rmdefault}{\mddefault}{\updefault}
$\E{6}$
}}}
\put(7276,-4411){\makebox(0,0)[lb]{\smash{\SetFigFont{12}{14.4}{\rmdefault}{\mddefault}{\updefault}
$\E{8}$
}}}
\put(8926,-2536){\makebox(0,0)[lb]{\smash{\SetFigFont{12}{14.4}{\rmdefault}{\mddefault}{\updefault}
$\fZ_{0s_2s_3}$
}}}
\put(8401,-2461){\makebox(0,0)[lb]{\smash{\SetFigFont{12}{14.4}{\rmdefault}{\mddefault}{\updefault}
$\E{6}$
}}}
\put(4951,-2011){\makebox(0,0)[lb]{\smash{\SetFigFont{12}{14.4}{\rmdefault}{\mddefault}{\updefault}
$v +a_1 \simeq \mbox{-} \frac13(\mbox{-}2, 1^2, 0^5| 0^8)$
}}}
\put(5401,-2536){\makebox(0,0)[lb]{\smash{\SetFigFont{12}{14.4}{\rmdefault}{\mddefault}{\updefault}
$\fZ_{1s_2s_3}$
}}}
\put(6376,-5086){\makebox(0,0)[lb]{\smash{\SetFigFont{12}{14.4}{\rmdefault}{\mddefault}{\updefault}
$v + 2 a_1 \simeq (0^8| 0^8)$
}}}
\put(7726,-4636){\makebox(0,0)[lb]{\smash{\SetFigFont{12}{14.4}{\rmdefault}{\mddefault}{\updefault}
$\fZ_{2s_2s_3}$
}}}
\end{picture}
}}}
}
As at the fixed points $\fZ_{2s_2s_3}$ of the first $T^2$ the local shift is
trivial,  the full spectrum there is equivalent to the $\E{8}$ model. 
The shift $v_{0s_2s_3}$ is equal to the shift of the $\E{6}$ model,
while the shift $v_{1s_2s_3} = - v_{0s_2s_3}$ of the fixed points
$\fZ_{1s_2s_3}$ gives rise to the complex conjugate states w.r.t.\  those
arising on the fixed point $\fZ_{0s_2s_3}$. For the twisted states one may
use table \ref{tab:z3models} to arrive at the spectrum given in the
introduction in \eqref{twistedstatesv=a}. We have used the schematic
picture, defined in figure \ref{fig:schemaOrbi} to give an overview of
the fixed point equivalent models that are induced at the fixed points
of the orbifold. 

In the reasoning presented in the introduction, we only took the four
dimensional zero modes into account. The twisted states zero
modes are of course localized at the fixed points, but the
untwisted states analysis was too naive in the sense that the collective
effects of the massive states of the untwisted sector from the
effective four dimensional viewpoint were ignored. These states were
taken into account in the direct calculation of the gaugino anomaly
presented in section \ref{sect:GauginoAnomaly}. 

In terms of figure \ref{fig:HetOverview} of the introduction the
problem may be stated as follows. As argued in section
\ref{sect:twistedhet} the zero mode limit of the twisted states does
not remove any states, as the twisted states are already four
dimensional. Therefore, we can safely go into the opposite direction
of the zero mode limit indicated in that figure. For the four
dimensional zero mode untwisted matter, however, we are not allowed to
do this, since in the zero mode limit many states of the super
Yang--Mills theory coupled to supergravity are removed. The correct
way of identifying the untwisted matter states, responsible for the
localized gauge anomalies, was discussed in section
\ref{sect:untwistedFixedZero} leading to the conditions
\eqref{GaugeMatterFixed}. 

The apparent paradox
of the introduction is thus resolved by taking into account all chiral
states, twisted and untwisted, present at a given fixed point. This is
conveniently described by using the fixed point equivalent pure
orbifold models discussed in the previous subsection.

%
%
\subsection{Anomalous $\boldsymbol{\U{1}}$s in heterotic $\Intr_3$ orbifolds}
\labl{sect:anomU1}

In the previous subsections we have drawn the conclusion, that in heterotic
$\Intr_3$ orbifold models there are never (localized) non--Abelian
anomalies. This followed directly from the analysis of the fixed point
equivalent pure orbifold models, where this statement can be proven by
direct inspection. In this section we use the same logic to
investigate the situation of anomalous $\U{1}$s in heterotic orbifold
models. Therefore, we start by summarizing the
well--known status of anomalous $\U{1}$s for pure orbifold models. 
By employing the fixed point equivalent models, we can use this
information to understand the localized anomalous $\U{1}$s in orbifold
models with Wilson lines, and relate this to the four dimensional zero
mode situation.

\subsubsection*{Pure orbifold models with an anomalous $\boldsymbol{\U{1}}$}
\labl{sect:pureOrbiU1}

The eight possible pure $\Intr_3$ orbifold models have been summarized
in table \ref{tab:z3models} of section \ref{sect:pureOrbi}. This table
tells us, that only four of those models have $\U{1}$ factors in their
gauge groups: the $\E{7}$, $\E{7}'$ and $\SU{9}$, $\SU{9}'$ model. 
The latter two have only one
$\U{1}$ factor, while the former ones have two. The precise forms of these
$\U{1}$ generators have been identified in \eqref{U1factorGens}. 
\begin{table}
\[
\renewcommand{\arraystretch}{1.5}
\arry{|l | l |l|}{\hline 
\text{Group} &~~ \U{1}\ \text{generator} & ~~\U{1}^3\  
\text{and mixed gravitational anomalies} 
\\\hline
\E{7} & 
\arry{l}{
q_s = ( \,0,  \, 1^2, \,0^5 ~|~ 0^8)
\\
k_{q_s} = 2 {q_s}^2  = 4
\\[.5ex]
q_s' = (\,0^8 ~|~ 1, \,0^7)
\\
k_{q_s'} = 2 {q_s'}^2 = 2
}
&
\arry{l}
{
\tr_{\rep{L}_s} q_s^3 \,  = 
\frac 19 ( 56 (1)^3 + 1 (-2)^3 ) + 
( 14 (\mbox{\small $\frac 23$})^3 
+ 1 (\mbox{\small $-\frac 43$})^3 )
+ 3 (\mbox{\small $\frac 23$})^3 
= 8 
\\
\tr_{\rep{L}_s} q_s ~ = 
\frac 19 ( 56 (1)~ + 1 (-2) ~\, ) + 
( 14 (\mbox{\small $\frac 23$}) ~ 
+ 1 (\mbox{\small $-\frac 43$}) ~\, )
+ 3 (\mbox{\small $\frac 23$}) ~ 
= 16 
\\[.5ex]
\!\tr_{\rep{L}_s} {q_s'}^3 = 
\frac 19 ( 64 (\mbox{\small $\frac 12$})^3 + 14 (-1)^3 ) + 
( 14 (\mbox{\small $-\frac 13$})^3 
+ 1 (\mbox{\small $\frac 23$})^3 )
+ 3 (\mbox{\small $\frac 23$})^3 
= 0
\\
\!\tr_{\rep{L}_s} {q_s'}~\,  = 
\frac 19 ( 64 (\mbox{\small $\frac 12$})~ + 14 (-1) ~\, ) + 
( 14 (\mbox{\small $-\frac 13$})~ 
+ 1 (\mbox{\small $\frac 23$}) ~\, )
+ 3 (\mbox{\small $\frac 23$}) ~ 
= 0 
} 
\\\hline 
\SU{9} &
\arry{l}{
q_s' = ( \, 0^8 ~|~ 1, \,0^7)
\\
k_{q_s'} = 2 {q_s'}^2 = 2
}
&
\arry{l}
{
\tr_{\rep{L}_s} {q_s'}^3  = 
\frac 19 ( 64 (\mbox{\small $\frac 12$})^3 + 14(-1)^3 ) + 
9 (\mbox{\small $\frac 23$})^3 + 3 \cdot 0 = 2
\\
\tr_{\rep{L}_s}  {q_s'} ~\, = 
\frac 19 ( 64 (\mbox{\small $\frac 12$}) ~ + 14(-1) ~\, ) + 
9 (\mbox{\small $\frac 23$})~ + 3 \cdot 0 = 8 
}
\\\hline
}
\]
\caption{The $\U{1}^3$ and mixed gravitational anomaly traces 
are computed for the two pure orbifold models that contain $\U{1}$
factors. We have included the level $k_q$ with which 
\eqref{U1universality} can be checked straightforwardly.
}
\labl{tab:U1traces}
\end{table}

For the $\E{7}$ and $\SU{9}$ models and all three $\U{1}$s we
calculate the pure gauge and mixed gravitational anomaly locally at a
fixed point in table \ref{tab:U1traces} by computing  
\equ{
\tr_{\rep{L}_s} Q_s^3 = \Bigl(  
\frac 1{27} \, 3 \, \tr_{\rep{R}_s} +  \tr_{\rep{S}_s}
+ 3\, \tr_{\rep{T}_s} \Bigr) Q_s^3,
\qquad 
\tr_{\rep{L}_s} Q_s = \Bigl(
\frac 1{27} \, 3 \, \tr_{\rep{R}_s} +  \tr_{\rep{S}_s}  
+ 3\, \tr_{\rep{T}_s} \Bigr) Q_s.
\labl{localTrace}
}
As the combination of traces in brackets in these equations 
will appear often in the discussion below, we use $\tr_{\rep{L}_s}$ to
denote the local trace over all representations present at fixed point
$\fZ_{s}$. 
Here, we only focus on fermions as sources of possible anomalous
$\U{1}$ contribution, we investigate the role of the anti--symmetric
tensor later.   
The results are displayed in table \ref{tab:U1traces}, using the
notation for the $\U{1}$ generators introduced in
\eqref{U1factorGens}. We see that both models have one anomalous
$\U{1}$.  

Apart from the pure and mixed gravitational
anomalies there are also mixed non--Abelian anomalies. It turns out
that anomalous $\U{1}$s at the fixed points are universal in the sense
that the following relation holds \cite{Schellekens:1987xh}  
(see also the discussion in \cite{Kobayashi:1997pb}) 
\equ{
\frac 16\, \frac{1}{k_{q_s}}\tr_{\rep{L}_s} 
\bigr( Q_s^3 \bigr) = 
\frac 12 \sum_{a} Q_s({\rep{L}_s^{(a)}}) \, I_2(\rep{L}_s^{(a)})
= 
\frac{1}{48}\, \tr_{\rep{L_s}} \bigl( Q_s \bigr). 
\labl{U1universality}
}
The sum is over the irreducible representations $\rep{L}_s^{(a)}$
contained in the fixed point representation  
$\rep{L}_s = \frac 1{27} 3 \rep{R}_s + \rep{S}_s + 3 \rep{T}_s$, 
with the same multiplicity factors. 
The quadratic indices $I_2(\rep{L}_s^{(a)})$ are normalized w.r.t.\ the 
simple factors local gauge group $G_s$, and $Q_s({\rep{L}_s^{(a)}})$ 
is the $\U{1}$ charge of $\rep{L}_s^{(a)}$ 
Because of the inclusion of the level  
\(
k_{q_s} = 2q_{s}^2
\)
of $Q_s$ this formula is valid for any normalization of this local
$\U{1}$ generator.

\subsubsection*{Wilson lines and anomalous $\boldsymbol{\U{1}}$s}
\labl{sect:WilsonU1}

The universality relation \eqref{U1universality} is consistent with 
the localized pure and mixed $\U{1}$ anomalies, of the ten dimensional
gaugino  \eqref{GauginoAnomaly} and the twisted states
\eqref{TwistedAnomaly},  computed before:
\equ{
\cA_{\U{1}} = 
\frac{-1}{(2\gp)^3} 
\sum_s \int 
\Bigl\{
\text{str}_{\rep{L}_s} 
\Bigl[
\frac 16\,  \gL_1 F_1^2 + \frac 12 \, \gL_1 \tF^2  
\Bigr]_{\rep{Ad}_s}
+ \frac 1{48}\, 
\tr_{\rep{L}_s}\Bigl [ \gL_1 \Bigr]_{\rep{Ad}_s} \tr\, R^2 
\Bigr\}\, 
\gd(z - \fZ_s -\gG)  \d^6 z. 
}
Here we have used the solutions to the descent equations
\eqref{DescentEq}; and $\d^6 z$ denotes the volume form of the torus. 
The local trace $\tr_{\rep{L}_s}$ over all representations at fixed
point $\fZ_{s}$ has been defined in \eqref{localTrace}. We have
decomposed the field strength 
\(
F|_{\rep{Ad}_s} = F_1 |_{\rep{Ad}_s}  + \tF |_{\rep{Ad}_s}
\)
at fixed point $\fZ_s$ in a $\U{1}$ part $F_1$ and a non--Abelian part
$\tF$, and $\gL_1|_{\rep{Ad}_s}$ denotes the infinitesimal $\U{1}$
parameter.  

In the previous subsection we investigated the pure orbifold models
with anomalous $\U{1}$ factor groups. Applying the fixed point
equivalent model analysis introduced in the previous section, we
immediately conclude, that in a $\Intr_3$ orbifold model with Wilson
lines there is a local anomalous $\U{1}$ factor group present at fixed
point $\fZ_s$, if the model at that fixed point is equivalent to one of the 
pure orbifold models $\E{7}$, $\E{7}'$, $\SU{9}$ or $\SU{9}'$. 

Moreover, the generator of the anomalous $\U{1}$ factor is identified
by the results of table \ref{tab:U1traces}.   There the pure and mixed
gravitational anomalies are calculated for the generators  $q_s$, or
$q_s'$, given in \eqref{U1factorGens}, when the
fixed point equivalent model is $\E{7}$, $\SU{9}'$, or $\E{7}'$,
$\SU{9}$, respectively. 

Obviously, at each fixed point there is at most one anomalous
$\U{1}$. However, in the heterotic orbifold models as a whole, there
may be many different anomalous $\U{1}$ generators, corresponding to
anomalous $\U{1}$s at different fixed points. It is therefore possible
that a generator of an anomalous $\U{1},$ at a given fixed point, is
not anomalous at another fixed point, or is even part of a
non--Abelian factor.  

In the case of different anomalous $\U{1}$ factors at various fixed
points, it is not possible to define a set of linear combinations of their 
generators, such that only one linear combination is anomalous at all
anomalous fixed points simultaneously, while all perpendicular local
$\U{1}$ generators are anomaly free at all fixed points.

The next question we address is how to identify the anomalous $\U{1}$
in the zero mode theory, if present. To this end we use the relation
between the localized fixed point anomaly and the zero mode anomaly,
given in \eqref{LocalGlobalAnomaly}, for the mixed gravitational
gauge anomaly \eqref{localTrace}:
\equ{
\cA_{\U{1}\, grav, zero} \propto \sum_s 
\tr_{\rep{L}_s} (\gL)  = 
\sum_s 
\tr_{\rep{L}_s} (H_I) \gL^I. 
}
(Here we have restricted the gauge parameter $\gL$ to the Cartan
subalgebra, since all localized anomalous $\U{1}$s are contained in the
Cartan of $\E{8}\times \E{8}'$.) 
From this it follows that there can only be an anomalous $\U{1}$ at
the zero mode level, when there is at least one localized anomalous
$\U{1}$; otherwise all local traces $\tr_{\rep{L}_s}( H_I)$ vanish. 
The gauge parameters $\gL = \gL^I H_I$ can be decomposed 
\equ{
\gL_\parallel =  \sum_{I,J}~\gL^I ~
\frac{\dsp 
\sum_s \tr_{\rep{L}_s}(H_I)\,  \sum_{t}\tr_{\rep{L}_{t}} (H_J)}
 {\dsp \sum_{K} \bigl( \sum_s \tr_{\rep{L}_s}(H_K)\bigr)^2} ~ H_J, 
\qquad 
\gL_\perp = \gL - \gL_\parallel
\labl{zeroModeU1s}
}
into parallel and perpendicular components with respect to the anomaly 
\equ{
\cA_{\U{1}\, grav, zero} \propto \sum_s \tr_{\rep{L}_s} (\gL_\parallel), 
\qquad \tr_{\rep{L}_s} (\gL_\perp) = 0.
}
Hence at the zero mode level one can define one anomalous
$\U{1}$, such that all perpendicular $\U{1}$ generators are anomaly
free. This anomalous $\U{1}$ is of course canceled by the standard
Green--Schwarz mechanism \cite{Green:1984sg}; due to the resulting
quadratically divergent Fayet--Iliopoulos D--term 
\cite{Fischler:1981zk} this symmetry is
spontaneously broken \cite{Dine:1987xk,Atick:1987gy,Dine:1987gj}.

\subsection{Examples with localized $\boldsymbol{\U{1}}$s}
\labl{sect:ExampleU1}

\subsubsection*{Wilson line induced anomalous $\boldsymbol{\U{1}}$s}

We close this section with two examples. Contrary to our previous
example, discussed in the introduction and section \ref{sect:v=a}, we
choose the gauge shift of the double $\E{6}$ embedding and Wilson line
\equ{
v ~ =  \mbox{$\frac 13$} \!
\left(\mbox{-}2,~1^2,~0^2, ~ 0^3 ~~|~~ \mbox{-}2,~1^2,~0, ~0^4\right),
\qquad 
a_1 =\mbox{$\frac 13$} \!
\left(~0, ~0^2, ~1^2,~0^3 ~~|~~ ~0,~0^2,\mbox{-}2,~0^4\right),
\labl{U1shiftWilson}
}
such that $\U{1}$ factors arise. 
The fixed point equivalent pure orbifold models are easily identified
using the method exposed in this section, the corresponding local
shifts are  
\equ{
\hspace{-18ex}
\arry{l}{
v_{0s_2s_3} = v ~~~~ ~~~~ = \mbox{$\frac 13$} 
\left(\mbox{-}2,~1^2,~0^2, ~ 0^3 ~~|~~ \mbox{-}2,~1^2,~0, ~0^4\right),
\\[1ex] 
v_{1s_2s_3} = v+ ~\, a_1 = \mbox{$\frac 13$} 
\left(\mbox{-}2,~1^2,~1^2, ~ 0^3 ~~|~~ \mbox{-}2,~1^2,\mbox{-}2, ~0^4\right),
\\[1ex]
v_{2s_2s_3} = v + 2 a_1 = \mbox{$\frac 13$} 
\left(\mbox{-}2,~1^2,\mbox{-}1^2, ~ 0^3 ~~|~~ \mbox{-}2,~1^2, ~2, ~0^4\right).
}
\hspace{-10ex} 
\raisebox{-10ex}{\scalebox{0.8}{\mbox{\begin{picture}(0,0)%
\includegraphics{modelU1.pstex}%
\end{picture}%
\setlength{\unitlength}{2763sp}%
\begingroup\makeatletter\ifx\SetFigFont\undefined%
\gdef\SetFigFont#1#2#3#4#5{%
  \reset@font\fontsize{#1}{#2pt}%
  \fontfamily{#3}\fontseries{#4}\fontshape{#5}%
  \selectfont}%
\fi\endgroup%
\begin{picture}(5325,3267)(3676,-5086)
\put(3676,-2011){\makebox(0,0)[lb]{\smash{\SetFigFont{12}{14.4}{\rmdefault}{\mddefault}{\updefault}
$\frac{1}{3} (\mbox{-}2, 1^4, 0^3 | \mbox{-}2,0^7) \simeq v + 2 a_1$
}}}
\put(6151,-2461){\makebox(0,0)[lb]{\smash{\SetFigFont{12}{14.4}{\rmdefault}{\mddefault}{\updefault}
$\SU{9}$
}}}
\put(7126,-5086){\makebox(0,0)[lb]{\smash{\SetFigFont{12}{14.4}{\rmdefault}{\mddefault}{\updefault}
$v = \frac{1}{3} (\mbox{-}2, 1^2, 0^5 | \mbox{-}2, 1^2, 0^5)$
}}}
\put(7726,-4636){\makebox(0,0)[lb]{\smash{\SetFigFont{12}{14.4}{\rmdefault}{\mddefault}{\updefault}
$\fZ_{0s_2s_3}$
}}}
\put(5401,-2536){\makebox(0,0)[lb]{\smash{\SetFigFont{12}{14.4}{\rmdefault}{\mddefault}{\updefault}
$\fZ_{2s_2s_3}$
}}}
\put(9001,-2536){\makebox(0,0)[lb]{\smash{\SetFigFont{12}{14.4}{\rmdefault}{\mddefault}{\updefault}
$\fZ_{1s_2s_3}$
}}}
\put(8026,-2011){\makebox(0,0)[lb]{\smash{\SetFigFont{12}{14.4}{\rmdefault}{\mddefault}{\updefault}
$v +a_1 \simeq  \frac{1}{3} (\mbox{-}2, 1^4, 0^3 | \mbox{-}2, 0^7)$
}}}
\put(7276,-4336){\makebox(0,0)[lb]{\smash{\SetFigFont{12}{14.4}{\rmdefault}{\mddefault}{\updefault}
$\E{6}^2$
}}}
\put(8026,-2461){\makebox(0,0)[lb]{\smash{\SetFigFont{12}{14.4}{\rmdefault}{\mddefault}{\updefault}
$\SU{9}$
}}}
\end{picture}
}}}
\labl{U1localShifts}
}
To show that these local shifts are equivalent to the ones indicated
in the picture, one may count the number of zeros of a shift and use
table \ref{tab:identifyShifts}. Obviously, also in this case the
non--Abelian anomalies cancel. Although the local
gauge shifts $v_{1s_2s_3}$ and $v_{2s_2s_3}$ are equivalent, they are
definitely not equal (up to an overall sign). 
This fact implies that the embedding of
$\SU{9}$ and  $\SO{14}' \times \U{1}'$ in $\E{8}$ and $\E{8}'$,
respectively, at the fixed points $\fZ_{1s_2s_3}$ and 
$\fZ_{2s_2s_3}$ is different. This can have some important
consequences, as we will see throughout this example.  

The first place where this can be noticed, is in the identification of
the generators of the anomalous $\U{1}$s at the fixed points
$\fZ_{1s_2s_3}$ and  $\fZ_{2s_2s_3}$. 
The general arguments of section \ref{sect:anomU1} explained that the
candidates for anomalous $\U{1}$s are given by \eqref{U1factorGens}. 
This means that their generators can be read off from the local shifts
\eqref{U1localShifts}. Therefore, the charges of the anomalous
$\U{1}$s of the $\SU{9}$ models are with respect to the generators
(cf. table \ref{tab:z3models}) 
\equ{
q_1' = 3 q'_{1s_2s_3} = \bigl(~0^8 ~~|~~ 1^3, ~1, ~0^4\bigr), 
\qquad 
q_2' = 3 q'_{1s_2s_3} = \bigl(~0^8 ~~| ~~ 1^3, \mbox{-}1, ~0^4\bigr), 
}
at fixed points $\fZ_{1s_2s_3}$ and  $\fZ_{2s_2s_3}$, respectively. 
This is an intriguing situation: consider the fixed points
$\fZ_{1s_2s_3}$, for example. While $q_1'$ generates the anomalous
$\U{1}$ there, the perpendicular part of $q_2'$ with respect to $q_1'$ is
one of the generators of $\SO{14}'$. 

Therefore, contrary to our previous example of section \ref{sect:v=a},
here the unbroken gauge group at the zero mode level 
\equ{
G = \SU{6}\!\times\!\SU{3}\!\times\!\U{1}
~\times~
\SO{8}'\!\times\!\SU{3}'\!\times\!\U{1}_+'\!\times\!\U{1}_-'
\labl{U1gaugeGroup}
}
forms a true subgroup $G = \cap_s G_s$ of the local gauge groups $G_s$
at the fixed points. (A discussion of the $\U{1}$ factors can be found
below.) As usual the twisted zero mode states are the same as
the twisted states at the fixed points, in addition some untwisted zero mode matter
arises. Since the twisted states may now be charged under all the zero
mode $\U{1}$ factors, we present the full zero mode matter spectrum 
in table \ref{tab:U1zeroModeSpectrum}. 
Again it may be checked that there is no non--Abelian anomaly, as is
to be expected from the general analysis. 
\begin{table}
\[
\renewcommand{\arraystretch}{1.25}
\arry{|lll|}{\hline 
\text{States} & \text{Representation} & \text{Spectrum}
~~~~
( \SU{6}, \SU{3} )_q (\SO{8}, \SU{3})'_{q_+', q_-'}
\\\hline 
\text{untwisted} &(\rep{3}_H, \rep{R} ~~~~~) &  
(\crep{15},\crep{3})_0(\rep{1},\rep{1})'_{0,0}
+ (\rep{1},\rep{1})_0(\rep{8},\crep{3})'_{1,0}
+ (\rep{1},\rep{1})_0(\rep{1},\crep{3})'_{-2,0}
\\[1ex]
\text{twisted} & (\rep{1}_H,\rep{S}_{0s_2s_3}) &
(\rep{1},\rep{3})_0(\rep{1},\rep{3})'_{0,0}
\\  & (\rep{1}_H,\rep{S}_{1s_2s_3}) &
(\rep{6},\rep{1})_{\mbox{-}\frac 13}(\rep{1},\rep{1})'_{1,\frac 13}
+ (\rep{1},\rep{3})_{\frac 23}(\rep{1},\rep{1})'_{1,\frac 13}
\\  & (\rep{1}_H,\rep{S}_{2s_2s_3}) &
 (\rep{6},\rep{1})_{\frac 13}(\rep{1},\rep{1})'_{1,\mbox{-}\frac 13}
+ (\rep{1},\rep{3})_{\mbox{-}\frac 23}(\rep{1},\rep{1})'_{1,\mbox{-}\frac 13}
\\\hline 
}
\]
\caption{The zero mode matter representations of the model with a
shift and a Wilson line,  given in \eqref{U1shiftWilson}, charged under
the zero mode gauge group \eqref{U1gaugeGroup}.
}
\labl{tab:U1zeroModeSpectrum}
\end{table}

There are three $\U{1}$ factors at the zero mode level. From the local
shifts, it can be inferred that the $\U{1}$ factor in the first
$\E{8}$ corresponds to the Cartan element 
\equ{
q= \bigl( ~0^3, ~1^2, ~0^3 ~~|~~ 0^8 \bigl).
}
Since the local shift $v_{0s_2s_3}$ of the fixed points $\fZ_{0s_2s_3}$
does not have entries in this direction, it follows that twisted
states at that fixed point should have net charge zero. As there is
just one irreducible representation at that fixed point, see table 
\ref{tab:z3models}, it follows that this charge is zero as indicated
in table \ref{tab:U1zeroModeSpectrum}. Moreover, since the shift
vectors $v_{1s_2s_3}$ and $v_{2s_2s_3}$ have opposite entries in the
direction of $q$, it follows that the $q$--charges of the twisted
states at fixed points  $\fZ_{1s_2s_3}$ and  $\fZ_{2s_2s_3}$, are
opposite. This is again, in agreement with the twisted spectrum of table
\ref{tab:U1zeroModeSpectrum}.  It can be confirmed that this $\U{1}$ 
is not anomalous.  The reason for this has been discussed in section
\ref{sect:nonAbelAnom}: at all fixed points this $\U{1}$ generator is
part of the set of generators of non--Abelian subgroups of $\E{8}$, 
see the picture in \eqref{U1localShifts}.

The two $\U{1}$ generators in the $\E{8}'$ are orthogonal linear
combinations of the anomalous generators $q_1'$ and $q_2'$ at the two
sets of fixed points  
\equ{
q_+' = \frac 12 (q_1' + q_2') = \bigl(~0^8 ~~|~~ 1^3, ~0, ~0^4 \bigr),
\qquad 
q_-' = \frac 12  (q_1' - q_2') = \bigl(~0^8 ~~|~~  0^3, ~1,~ 0^4 \bigr),
}
such that only $q_+'$ is anomalous. This is a special case of the 
general results given in \eqref{zeroModeU1s}. Also, in this specific
example, it is not possible to define one linear combination of these 
$\U{1}$ generators which is responsible for both the localized and
global anomalous $\U{1}$s at the same time.

\subsubsection*{All non--trivial equivalent models at the fixed points}

\begin{table}[ht]
{\small 
\[
\renewcommand{\arraystretch}{1.25}
\arry{|lll|}{\hline 
\text{States} & \text{Representation} & \text{Spectrum} 
~~~~
(\SU{6}, \SU{3})_q (\SO{8}, \SU{3})'_{q_-', q_+'}
\\\hline 
\text{untwisted} &(\rep{3}_H, \rep{R} ~~~~\,) &  
(\mathbf{\bar{3}},\mathbf{\bar{6}})_{\mbox{-}2}(\mathbf{1},\mathbf{1})'_{0,0}
\\[1ex]
\text{twisted} & (\rep{1}_H,\rep{S}_{00s_3}\,) &
(\mathbf{1},\mathbf{15})_{0}(\mathbf{1},\mathbf{1})'_{0,0} + 
(\mathbf{1},\mathbf{\bar{6}})_{\mbox{-}2}(\mathbf{1},\mathbf{1})'_{0,0} + 
(\mathbf{1},\mathbf{\bar{6}})_{2}(\mathbf{1},\mathbf{1})'_{0,0} 
\\ & (\rep{3}_H,\rep{T}_{00s_3}) &
(\mathbf{3},\mathbf{1})_{0}(\mathbf{1},\mathbf{1})'_{0,0}
\\ & (\rep{1}_H,\rep{S}_{10s_3}\,) &
(\rep{1},\rep{3})_0(\rep{1},\rep{3})'_{0,0}
\\ & (\rep{1}_H,\rep{S}_{20s_3}\,) &
(\rep{1},\rep{3})_0(\rep{1},\crep{3})'_{0,0}
\\[2ex] 
& (\rep{1}_H,\rep{S}_{01s_3}\,) &
(\mathbf{1},\mathbf{1})_{\mbox{-}\frac{4}{3}}(\mathbf{8},\mathbf{1})'_{0,\mbox{-}\frac{1}{3}}+
(\mathbf{1},\mathbf{1})_{\mbox{-}\frac{4}{3}}(\mathbf{1},\mathbf{3})'_{\mbox{-}1,\mbox{-}\frac{1}{3}}+  
(\mathbf{1},\mathbf{1})_{\mbox{-}\frac{4}{3}}(\mathbf{1},\mathbf{\bar{3}})'_{1,\mbox{-}\frac{1}{3}}+
(\mathbf{1},\mathbf{1})_{\frac{8}{3}}(\mathbf{1},\mathbf{1})'_{0,\frac{2}{3}}
\\ & (\rep{3}_H,\rep{T}_{01s_3}) &
(\mathbf{1},\mathbf{1})_{\mbox{-}\frac{4}{3}}(\mathbf{1},\mathbf{1})'_{0,\frac{2}{3}}
\\  & (\rep{1}_H,\rep{S}_{11s_3}\,) &
(\mathbf{1},\mathbf{1})_{\mbox{-}\frac{4}{3}}(\mathbf{8},\mathbf{1})'_{\mbox{-}\frac{1}{2},\frac{1}{6}}+  
(\mathbf{1},\mathbf{1})_{\mbox{-}\frac{4}{3}}(\mathbf{1},\mathbf{3})'_{0,\frac{2}{3}}+ 
(\mathbf{1},\mathbf{1})_{\mbox{-}\frac{4}{3}}(\mathbf{1},\mathbf{\bar{3}})'_{\mbox{-}1,\mbox{-}\frac{1}{3}}+ 
(\mathbf{1},\mathbf{1})_{\frac{8}{3}}(\mathbf{1},\mathbf{1})'_{1,\mbox{-}\frac{1}{3}}
\\ & (\rep{3}_H,\rep{T}_{11s_3}) &
(\mathbf{1},\mathbf{1})_{\mbox{-}\frac{4}{3}}(\mathbf{1},\mathbf{1})'_{1,\mbox{-}\frac{1}{3}}
\\  & (\rep{1}_H,\rep{S}_{21s_3}\,) &
(\mathbf{1},\mathbf{1})_{\mbox{-}\frac{4}{3}}(\mathbf{8},\mathbf{1})'_{\frac{1}{2},\frac{1}{6}}+ 
(\mathbf{1},\mathbf{1})_{\mbox{-}\frac{4}{3}}(\mathbf{1},\mathbf{\bar{3}})'_{0,\frac{2}{3}}+ 
(\mathbf{1},\mathbf{1})_{\mbox{-}\frac{4}{3}}(\mathbf{1},\mathbf{3})'_{1,\mbox{-}\frac{1}{3}}+
(\mathbf{1},\mathbf{1})_{\frac{8}{3}}(\mathbf{1},\mathbf{1})'_{\mbox{-}1,\mbox{-}\frac{1}{3}}
\\ & (\rep{3}_H,\rep{T}_{21s_3}) &
(\mathbf{1},\mathbf{1})_{\mbox{-}\frac{4}{3}}(\mathbf{1},\mathbf{1})'_{\mbox{-}1,\mbox{-}\frac{1}{3}}
\\[2ex]
  & (\rep{1}_H,\rep{S}_{02s_3}\,) &
(\mathbf{1},\mathbf{6})_{\mbox{-}\frac{2}{3}}(\mathbf{1},\mathbf{1})'_{0,\mbox{-}\frac{2}{3}}+ 
 (\mathbf{\bar{3}},\mathbf{1})_{\frac{4}{3}}(\mathbf{1},\mathbf{1})'_{0,\mbox{-}\frac{2}{3}}
\\  & (\rep{1}_H,\rep{S}_{12s_3}\,) &
(\mathbf{1},\mathbf{6})_{\mbox{-}\frac{2}{3}}(\mathbf{1},\mathbf{1})'_{1,\frac{1}{3}}+ 
(\mathbf{\bar{3}},\mathbf{1})_{\frac{4}{3}}(\mathbf{1},\mathbf{1})'_{1,\frac{1}{3}}
\\  & (\rep{1}_H,\rep{S}_{22s_3}\,) &
(\mathbf{1},\mathbf{6})_{\mbox{-}\frac{2}{3}}(\mathbf{1},\mathbf{1})'_{\mbox{-}1,\frac{1}{3}}+
(\mathbf{\bar{3}},\mathbf{1})_{\frac{4}{3}}(\mathbf{1},\mathbf{1})'_{\mbox{-}1,\frac{1}{3}}
\\\hline 
}
\]}
\caption{The zero mode matter representations of the model with a
shift and two Wilson lines,  given in \eqref{AllshiftWilson}, charged under
the zero mode gauge group \eqref{AllgaugeGroup}.
}
\labl{tab:AllzeroModeSpectrum}
\end{table}

\equ{
\hspace{-10ex}
\arry{l}{
 v ~ = \mbox{$\frac 13$} 
\left(\mbox{-}2,~\,1^2,~0, ~0^4 ~~|~~ ~0,~0^2, ~0, ~0^4\right),
\\[1ex] 
 a_1 = \mbox{$\frac 13$} 
\left(~0,~0^2,\mbox{-}2, ~1^4 ~~|~~ ~0,~0^2,\mbox{-}2, ~0^4\right),
\\[1ex]
 a_2 = \mbox{$\frac 13$} 
\left(~0,~0^2, ~0, ~0^4 ~~|~~ \mbox{-}2,~1^2, ~0, ~0^4\right).
}
\raisebox{-10ex}{\scalebox{0.8}{\mbox{\begin{picture}(0,0)%
\includegraphics{modelall.pstex}%
\end{picture}%
\setlength{\unitlength}{2763sp}%
\begingroup\makeatletter\ifx\SetFigFont\undefined%
\gdef\SetFigFont#1#2#3#4#5{%
  \reset@font\fontsize{#1}{#2pt}%
  \fontfamily{#3}\fontseries{#4}\fontshape{#5}%
  \selectfont}%
\fi\endgroup%
\begin{picture}(6515,6000)(-299,-8836)
\put(1726,-5236){\makebox(0,0)[lb]{\smash{\SetFigFont{12}{14.4}{\rmdefault}{\mddefault}{\updefault}
$\SU{9}$
}}}
\put(4351,-4786){\makebox(0,0)[lb]{\smash{\SetFigFont{12}{14.4}{\rmdefault}{\mddefault}{\updefault}
$\fZ_{10s_3}$
}}}
\put(4051,-5236){\makebox(0,0)[lb]{\smash{\SetFigFont{12}{14.4}{\rmdefault}{\mddefault}{\updefault}
$\E{7}$
}}}
\put(6151,-5236){\makebox(0,0)[lb]{\smash{\SetFigFont{12}{14.4}{\rmdefault}{\mddefault}{\updefault}
$\E{7}$
}}}
\put(5101,-3061){\makebox(0,0)[lb]{\smash{\SetFigFont{12}{14.4}{\rmdefault}{\mddefault}{\updefault}
$\E{7}$
}}}
\put(1951,-8836){\makebox(0,0)[lb]{\smash{\SetFigFont{12}{14.4}{\rmdefault}{\mddefault}{\updefault}
$\E{6}$
}}}
\put(4051,-8836){\makebox(0,0)[lb]{\smash{\SetFigFont{12}{14.4}{\rmdefault}{\mddefault}{\updefault}
$\E{6}^2$
}}}
\put(3001,-6661){\makebox(0,0)[lb]{\smash{\SetFigFont{12}{14.4}{\rmdefault}{\mddefault}{\updefault}
$\E{6}^2$
}}}
\put(3451,-8386){\makebox(0,0)[lb]{\smash{\SetFigFont{12}{14.4}{\rmdefault}{\mddefault}{\updefault}
$\fZ_{01s_3}$
}}}
\put(1351,-4786){\makebox(0,0)[lb]{\smash{\SetFigFont{12}{14.4}{\rmdefault}{\mddefault}{\updefault}
$\fZ_{21s_3}$
}}}
\put(5551,-4786){\makebox(0,0)[lb]{\smash{\SetFigFont{12}{14.4}{\rmdefault}{\mddefault}{\updefault}
$\fZ_{11s_3}$
}}}
\put(751,-3736){\makebox(0,0)[lb]{\smash{\SetFigFont{12}{14.4}{\rmdefault}{\mddefault}{\updefault}
$\fZ_{22s_3}$
}}}
\put(4951,-3736){\makebox(0,0)[lb]{\smash{\SetFigFont{12}{14.4}{\rmdefault}{\mddefault}{\updefault}
$\fZ_{12s_3}$
}}}
\put(2851,-7336){\makebox(0,0)[lb]{\smash{\SetFigFont{12}{14.4}{\rmdefault}{\mddefault}{\updefault}
$\fZ_{02s_3}$
}}}
\put(2251,-8386){\makebox(0,0)[lb]{\smash{\SetFigFont{12}{14.4}{\rmdefault}{\mddefault}{\updefault}
$\fZ_{00s_3}$
}}}
\put(676,-3061){\makebox(0,0)[lb]{\smash{\SetFigFont{12}{14.4}{\rmdefault}{\mddefault}{\updefault}
$\SU{9}$
}}}
\put(151,-4786){\makebox(0,0)[lb]{\smash{\SetFigFont{12}{14.4}{\rmdefault}{\mddefault}{\updefault}
$\fZ_{20s_3}$
}}}
\put(-299,-5236){\makebox(0,0)[lb]{\smash{\SetFigFont{12}{14.4}{\rmdefault}{\mddefault}{\updefault}
$\SU{9}$
}}}
\end{picture}
}}}
\labl{AllshiftWilson}
}
Our final example contains two Wilson lines in addition to the gauge
shift given in the equation \eqref{AllshiftWilson}, above. This model
thus provides an example of the second diagram in figure
\ref{fig:schemaOrbi}.  

This model
has the feature that all non--trivial fixed point equivalent models of
table \ref{tab:z3models} appear at one or more fixed points, as can be
seen from the local shift vectors
\equ{
\arry{l}{
 v_{00s_3} = \mbox{$\frac 13$} 
\left(\mbox{-}2,~\,1^2,~0, ~0^4 ~~|~~ ~0,~0^2, ~0, ~0^4\right),
\\[1ex] 
v_{10s_3}  = \mbox{$\frac 13$} 
\left(\mbox{-}2,~\,1^2, ~0, ~0^4 ~~|~~ \mbox{-}2,~1^2, ~0, ~0^4\right),
\\[1ex]
v_{20s_3}  = \mbox{$\frac 13$} 
\left(\mbox{-}2,~\,1^2, ~0, ~0^4 ~~|~~ ~2,\mbox{-}1^2, ~0, ~0^4\right),
\\[1ex]
v_{01s_3} = \mbox{$\frac 13$} 
\left(\mbox{-}2,~\,1^2,\mbox{-}2, ~1^4 ~~|~~ ~0,~0^2, \mbox{-}2, ~0^4\right),
\\[1ex] 
v_{11s_3}  = \mbox{$\frac 13$} 
\left(\mbox{-}2,~\,1^2, \mbox{-2}, ~1^4 ~~|~~ \mbox{-}2,~1^2, \mbox{-}2, ~0^4\right),
}
\quad 
\arry{l}{
v_{21s_3} = \mbox{$\frac 13$} 
\left(\mbox{-}2,~\,1^2, \mbox{-}2, ~1^4 ~~|~~ ~2, \mbox{-}1^2, \mbox{-}2, ~0^4\right),
\\[1ex]
v_{02s_3} = \mbox{$\frac 13$} 
\left(\mbox{-}2,~\,1^2,~2, \mbox{-}1^4 ~~|~~ ~0,~0^2, ~2, ~0^4\right),
\\[1ex] 
v_{12s_3}  = \mbox{$\frac 13$} 
\left(\mbox{-}2,~\,1^2, ~2, \mbox{-}1^4 ~~|~~ \mbox{-}2,~1^2, ~2, ~0^4\right),
\\[1ex]
v_{22s_3} = \mbox{$\frac 13$} 
\left( \mbox{-}2,~\,1^2,~2, \mbox{-}1^4 ~~|~~ ~2,\mbox{-}1^2, ~2, ~0^4\right).
\\[1ex]\\[1ex]
}
}
Using these shifts and table \ref{tab:identifyShifts}, we composed the
figure in \eqref{AllshiftWilson}. Since all non--trivial fixed point
models arise at the fixed points, both the $\E{7}$ and $\SU{9}$
equivalent models give rise to anomalous $\U{1}$s at the fixed points. 
These generators take the form 
\equ{
\arry{l}{
q_{01s_3} = \bigl( ~ 1^8 ~~|~~ 0^8 \bigr),
\\[1ex]
q_{11s_3} = \bigl( ~ 1^8 ~~|~~ 0^8 \bigr),
\\[1ex]
q_{21s_3} = \bigl( ~ 1^8 ~~|~~ 0^8 \bigr),
}
\qquad 
\arry{l}{
q_{01s_3} = \mbox{-}q_{02s_3} = \bigl( ~ 0^8 ~~|~~ ~0^3, ~1, 0^4 \bigr),
\\[1ex]
q_{11s_3} = \mbox{-}q_{22s_3} = \bigl( ~ 0^8 ~~|~~ ~1^3, ~1, 0^4 \bigr),
\\[1ex]
q_{21s_3} = \mbox{-}q_{12s_3} =  \bigl( ~ 0^8 ~~|~~ \mbox{-}1^3, ~1, 0^4 \bigr).
}
}
Therefore, it is in this case even more complicated to see directly,
which linear combination of anomalous $\U{1}$ generators at the fixed
points, is the generator of the anomalous $\U{1}$ at the zero mode
level. In the zero mode gauge group  
\equ{
G = \SU{6}\!\times\!\SU{3}\!\times\!\U{1}
~\times~
\SO{8}'\!\times\!\SU{3}'\!\times\!\U{1}_+'\!\times\!\U{1}_-'
\labl{AllgaugeGroup}
}
we find again three $\U{1}$ factors; two of them in the 
$\E{8}'$ group. The corresponding shift of these $\U{1}$s are given by 
\equ{
 q  = \mbox{$\frac 13$} 
\left( ~1^8 ~~|~~ ~0,~0^2, ~0, ~0^4\right),
\qquad 
 \arry{l}{
 q_+' = \mbox{$\frac 13$} 
\left(~0^8 ~~|~~ ~0^3, ~1, ~0^4\right),
\\[1ex]
q_-' = \mbox{$\frac 13$} 
\left(~0^8 ~~|~~ ~1^3, ~0, ~0^4\right).
}
\labl{AllzeroU1}
}
The zero mode matter spectrum is collected in table
\ref{tab:AllzeroModeSpectrum}. Using \eqref{zeroModeU1s} 
we identify the zero mode anomalous $\U{1}$ to be generated by 
$q$.

%
%
\section{Conclusions and outlook}
\labl{sect:conclusion}

We have investigated the structure of local anomalies of heterotic
orbifold models. For this purpose we considered the field theory limit of
the heterotic $\E{8}\times \E{8}'$ string compactified on the orbifold
$T^6/\Intr_3$. The main results of this endeavor can be summarized as
follows:
\begin{list}{}{
\setlength{\leftmargin}{2ex}
\setlength{\labelsep}{5ex}
\setlength{\parsep}{0ex}
}
\item
We have computed the gaugino anomaly 
in the presence of (shift embedded) Wilson lines that commute with the
orbifold boundary conditions, using the Fujikawa method. 
The result of this calculation
\eqref{GauginoAnomaly} shows that the anomaly becomes localized at the
orbifold fixed points, and depends crucially on the local
spectra of untwisted states \eqref{GaugeMatterFixed}. 
In addition there is a ten dimensional anomaly on the orbifold, which
has a normalization factor of $1/3$ compared to the anomaly on the torus.
\item
Combining this result with the twisted spectra obtained from string
theory, it followed, that there are no non--Abelian anomalies in heterotic 
$\Intr_3$ orbifold models with shift embedded Wilson lines. 
\item
This conclusion can be drawn by direct inspection for all possible
models with non--vanishing Wilson lines. However, to
understand the local structure of orbifold models with Wilson lines
better, we employed the notion of fixed point equivalent models. 
Two models are said to be fixed point equivalent, if at a given fixed
point their defining gauge shifts are equal up to Weyl reflections and
lattice shifts. This implies that at those fixed points their twisted
and untwisted spectra are isomorphic. 
\\
It followed, that any orbifold model of this class at a given fixed
point is equivalent to one of the eight
pure orbifold models, which are summarized in table \ref{tab:z3models}. 
As none of them suffers from a non--Abelian anomaly, the conclusion
of the previous point is confirmed.  
\item 
By using fixed point equivalent models, it is easy to show that at each
fixed point there is at most one anomalous $\U{1}$ present, which exists
only if its fixed point equivalent model is either
 $\E{7}$, $\E{7}'$, $\SU{9}$ or $\SU{9}'$ (see table
\ref{tab:z3models}).  However, in general the
anomalous $\U{1}$s at the different fixed points correspond to
different generators of the Cartan subgroup of
$\E{8}\times\E{8}'$. Because of this, it is not possible to define a
single linear combination which is 
the sole source of the $\U{1}$ anomalies, at the various fixed points.
At the effective four dimensional zero mode level a single anomalous
$\U{1}$ can be defined from the local anomalous $\U{1}$ generators. 
\end{list}
Let us conclude by giving an outlook on possible future
directions based on the work presented in this paper:
\begin{list}{}{
\setlength{\leftmargin}{2ex}
\setlength{\labelsep}{5ex}
\setlength{\parsep}{0ex}
}
\item 
The method we employed to compute the gaugino gauge anomaly can be
applied to a much wider range of calculations on orbifolds. We have used
an explicit orbifold projection operator in our computation, to be
able to use the mode functions of the torus. This avoids having to work
with the more complicated mode functions on the orbifold. It may be
checked that for computations on five dimensional orbifolds this
technique is also very powerful. 
\item 
Throughout this paper we have restricted ourselves to heterotic
$\E{8}\times\E{8}'$ string theory on the simplest six dimensional
prime orbifold $T^6/\Intr_3$. The calculation of the gaugino anomaly,
for example, can be extended to other prime or non--prime orbifolds,
of six or other dimensions. This leads to interesting questions, 
for example, whether non--prime orbifold models could have a more
complicated anomaly structure than the one considered here? 
\item 
Also the convenient concept of fixed point equivalent models will definitely
have useful applications to other (heterotic) orbifold models. In
addition, this tool can be useful for more phenomenological
applications, as it gives insight in the structure of the theory at the
fixed points, and its zero modes.
\item 
We showed, that the structure of localized anomalous
$\U{1}$s in heterotic orbifolds can be quite
complicated. Therefore, we will investigate the localized
Fayet--Iliopoulos terms for these models in a future publication. 
In particular we would like to settle the question, whether similar
destabilization effects are at work as in supersymmetric five
dimensional orbifold models
\cite{GrootNibbelink:2002qp,GrootNibbelink:2002wv}. 
\end{list}

\section*{Acknowledgments}

We would like to thank T.\ Kobayashi, W.\ Buchm\"uller, 
L.\ Covi for useful discussions. 
\\ 
Work supported in part by the European Community's Human Potential
Programme under contracts HPRN--CT--2000--00131 Quantum Spacetime,
HPRN--CT--2000--00148 Physics Across the Present Energy Frontier
and HPRN--CT--2000--00152 Supersymmetry and the Early Universe.
SGN was supported by priority grant 1096 of the Deutsche
Forschungsgemeinschaft. 
MO was partially supported by the Polish KBN grant 2 P03B 052 16.

\appendix

%
%

\section{Spinors in relevant dimensions}
\labl{SpinorDims}

In this appendix we provide some of the details involving the
decomposition of the $D=(1,9)$ spinor representation when compactified
on a six dimensional torus or orbifold. Many details are omitted, they can
be found in \cite{VanProeyen:1999ni,pol_2}; here
we have focused on the material needed for proper understanding of the
discussions in the main text. 
Because of the orbifold twist action \eqref{OrbiTwistCoor}, we further
decompose the  spinor representation on $T^6$ into $T^2$
subrepresentations. As all information concerning the spinor
representation is encoded in the Clifford algebra, we start by
describing the Clifford algebras in $D = 2; 6; (1,3);$ and $(1, 9)$
dimensions.  

The following properties hold in any of these dimensions. We described
them with generic generators $\gG_M$ of a $D$--dimensional Clifford
algebra. The generators, the chiral operator $\tgG$ and the Dirac
conjugation of a fermion $\gps$ are defined by 
\equ{
\{ \gG_M, \gG_N \} = 2 \get_{MN} \Id, 
\quad 
\tgG = \ga_D \prod \gG_M, 
\quad 
\gG_M^\dag =\begin{cases}
\gG_M & \text{Eucl.}, \\
\gG_0 \gG_M \gG_0 & \text{Mink.}
\end{cases} 
\quad 
\bgps = \begin{cases}
\gps^\dag & \text{Eucl.}, \\
\gps^\dag \gG_0 & \text{Mink.}
\end{cases}, 
}
where $\get_{MN}$ denotes a Minkowski or Euclidean metric with 
signatures $(-1, 1, \ldots)$ or $(1, 1, \ldots)$, respectively. 
(The factor $\ga_D$ is chosen such that the chirality operator $\tgG$
is Hermitian.) In addition in even dimensions there are two charge
conjugation matrices $S_\pm$ with the properties
\equ{
S_\pm\inv \gG_M S_\pm =  \pm \gG_M^T, 
\qquad 
S_\pm^\dag = S_\pm\inv = S_\pm, 
\qquad 
\gps^{S_{\pm}} = S_\pm \bgps^T,
}
and $\gps^{S_\pm}$ denotes the Majorana conjugations with respect to
both charge conjugation operators. 
In the table \ref{CliffordDim} we give an explicit basis representation of the
Clifford algebra generators, the chirality operator and the charge
conjugation operators, and give their dimension dependent properties. 

\begin{table}
\equ{ 
\renewcommand{\arraystretch}{1.5}
\arry{ l | c | c | c | c }{
\text{Dimension} & 2 & 6 & (1,3) & (1,9) 
\\ \hline 
\text{Generators} & 
\arry{l}{
\gs_1 = \mbox{\scriptsize $\pmtrx{0 & ~1\\ 1 & ~0}$} \\[5ex] 
\gs_2 = \mbox{\scriptsize $\pmtrx{0 & \mbox{-}i\\ i & ~0}$}
}
& 
\arry{l}{
\tgs_1 = ~\Id \otimes ~\Id \otimes \gs_1 \\ 
\tgs_2 = ~\Id \otimes ~\Id \otimes \gs_2 \\ 
\tgs_3 = ~\Id \otimes \gs_1 \otimes \gs_3 \\ 
\tgs_4 = ~\Id \otimes \gs_2 \otimes \gs_3 \\ 
\tgs_5 = \gs_1 \otimes \gs_3 \otimes \gs_3 \\ 
\tgs_6 = \gs_2 \otimes \gs_3 \otimes \gs_3 
}
& 
\arry{l}{
\gg_0 = i \Id \otimes \gs_1 \\ 
\gg_1 = ~ \Id \otimes \gs_2 \\ 
\gg_2 =  \gs_1 \otimes \gs_3 \\ 
\gg_3 =  \gs_2 \otimes \gs_3 
}
&
\arry{l}{
\gG_\gm = \Id_6 \otimes \gg_\gm \\ 
\gG_{a+3} = \tgs_a \otimes \tgg 
}
\\ \hline 
\text{Chirality} & 
\gs_3 = \mbox{\scriptsize $\pmtrx{1 & ~0 \\ 0 & \mbox{-}1}$} 
& 
\tgs~ = \gs_3 \otimes \gs_3 \otimes \gs_3 
& 
\tgg = \gs_3 \otimes \gs_3 
& 
\tgG = \tgs \otimes \tgg
\\ \hline 
\text{Charge conj.} & 
\arry{l}{
s_+ = \gs_1, ~  
s_- = \gs_2 
}
& 
\arry{l}{
c_+ = s_+ \otimes s_- \otimes s_+ \\ 
c_- = s_- \otimes s_+ \otimes s_- 
}
& 
\arry{l}{
C_+ = s_- \otimes s_+ \\
C_- = s_+ \otimes s_-
}
&
S_\pm = c_\pm \otimes C_\pm 
\\
& 
s_\pm^T = \pm s_\pm 
& 
c_\pm^T = \mp c_\pm 
& 
C_\pm^T = - C_\pm 
& 
S_\pm^T = \pm S_\pm 
\\
& 
s_\pm\inv \gs_3 s_\pm = - \gs_3^T
&
c_\pm\inv \tgs c_\pm = -\tgs^T
&
C_\pm\inv \tgg C_\pm = +\tgg^T
&
S_\pm\inv \tgG S_\pm = - \tgG^T
\\ \hline 
\text{Majorana cond.} &
\gps^{\gs_+} = \gps
&
 \gps^{c_-} = \gps
&
 \gps^{C_-} = \gps
&
 \gps^{S_-} = \gps
\\ \hline 
\text{Majorana-Weyl} & 
\text{no} & \text{no} & \text{no} & \text{yes} 
}
\non}
\caption{This table summarizes the dimension dependent properties of
the Clifford algebra in the dimensions $D = 2; 6; (1,3); (1,9)$. Using
tensor products an explicit basis is indicated for the generators of
the Clifford algebra, the chirality operator and the charge
conjugation matrices. Furthermore, we have indicated which charge
conjugation can be used to construct Majorana fermions; only in 
$D = (1,9)$ a spinor can be both Majorana and Weyl.}
\labl{CliffordDim}
\end{table}

Next we exploit the properties indicated in table \ref{CliffordDim} to
discuss how a generic $D=(1,9)$ dimensional Majorana-Weyl spinor 
$\gps$ with chirality $\gb = \pm$ can be decomposed. 
Let $\gx^{\ga}$ denote a spinor in two dimensions with chirality 
$\ga = \pm$. By taking a tensor product 
of three of such spinors, a chiral spinor 
$\get^\ga = \get^{\ga_1 \ga_2 \ga_3}$ 
in six dimensions is obtained, which satisfies 
\equ{
-i \gS_i \, \get^\ga = \ga_i\, \get^\ga,
\qquad \tgs\, \get^\ga = (-)^\ga \, \get^\ga = \ga_1 \ga_2 \ga_3\, \get^\ga.
\labl{TorusChirality}
}
The Majorana-Weyl spinor can then be decomposed as 
\equ{
\gps = \frac 1{\sqrt 2} \sum_{\ga} 
\get^{\ga_1 \ga_2 \ga_3} \otimes \gps^{\ga_1 \ga_2 \ga_3}
= \frac 1{\sqrt 2} \sum_{\ga} 
\gx^{\ga_1} \otimes \gx^{\ga_2} \otimes \gx^{\ga_3} \otimes 
\gps^{\ga_1 \ga_2 \ga_3}.
}
According to table \ref{CliffordDim} the $(1,9)$ dimensional chirality
and the charge conjugation matrices can be written in terms of the 
$2 \times 2 \times 2$ dimensional representations as 
\equ{
\tgG = \gs_3 \otimes \gs_3 \otimes \gs_3 \otimes \tgg, 
\qquad 
S_- = (- \gs_3 \otimes \Id \otimes \gs_3 \otimes \Id) 
(s_+ \otimes s_+ \otimes s_+ \otimes C_-), 
}
hence this gives the chiral and charge conjugation properties
of the $D=(1,3)$ dimensional spinors $\gps^{\ga_1 \ga_2 \ga_3}$: 
\equ{
\gb\, \ga_1 \ga_2 \ga_3 \, \tgg\,  \gps^{\ga_1 \ga_2 \ga_3} = 
(- \ga_1 \ga_3)  ( \gps^{-\!\ga_1 -\!\ga_2 -\!\ga_3})^{C_-} = 
\gps^{\ga_1 \ga_2 \ga_3}.
\labl{MWDecomp}
}
Now we choose a chiral basis for the fermions in $D=(1,3)$ dimensions
of chirality $\gb$, i.e.\ $ \ga_1 \ga_2 \ga_3 = \gb = \pm $, and see 
that the states with opposite chirality are related by the four
dimensional Majorana condition. The $D=(1,3)$ dimensional spinors
form a singlet and a triplet under the holonomy group $\SU{3}_H$:  
\equ{
\gps^0 = \gps^{+++}, \quad 
\gps^a = (\gps^{+--}, \gps^{-+-},  \gps^{--+}).
}
Notice that it is important to keep track of the ten dimensional
chirality, since not all fermions have the same chirality: the
gauginos $\gch$ and the gravitino $\gps_M$ are left--handed, 
while the dilatino $\gl$ is right--handed.

%
%
\section{Gaugino mode functions on $\boldsymbol{T^6}$ with Wilson lines} 
\labl{sect:GauginoWave}

This appendix is devoted to the construction of a complete set of
torus mode functions for scalars and  gauginos, which take the Wilson
lines into account. The calculation of the gaugino anomalies, presented in
section \ref{sect:GauginoAnomaly}, relies heavily on the material
developed here.  

The mode functions $\gf_q(z)$, of the torus defined above 
\eqref{OrbiTwistCoor}, are the periodic scalar functions on $\Cplx^3$
\equ{
\left.
\arry{l}{
\gf_q(z + ~~ \hat \imath) = \gf_q(z)
\\[1mm] 
\gf_q(z + \gth\, \hat \imath) = \gf_q(z) 
}\right\}
\quad \Ra \quad 
\gf_q(z) = N_q \, e^{2\pi i (q_i z^i + \bq_\ui \bz^\ui)/R_i},
\quad 
\pmtrx{ q_i \\ \bq_\ui} = 
\frac 1{\bgth - \gth} 
\pmtrx{ ~~\bgth n_i - m_i \\ - \gth n_i + m_i},
\labl{TorusWave}
}
with $n_i, m_i \in \Intr$. The normalization $N_q$ is chosen such that
these wave functions are orthonormal and form a complete set on the 
torus $T^6$
\equ{
\int_{T^6} \!\d z\,  \gf_q^\dag(z) \gf_{q'}(z) = \gd_{q\,  q'}, 
\qquad 
\sum_{q}  \gf_q(z) \gf_q^\dag(z') = \gd(z - z' - \gG).
\labl{TorusComplete}
}
The gaugino can be decomposed as 
\(
{\dsp 
\gch(x, z) = \frac 1{\sqrt 2} \sum_{\ga, A, q}
\get^\ga_{qA}(z) \, \gch^\ga_{qA}(x) 
}
\)
with the help of the mode functions 
\equ{
\arry{l}{
\get^\ga_{qA}(z) =  \gf_q(z) \, T(z) T_A T\inv(z) \, \get^\ga ~\,
= \gf_q(z) \, e^{~2\pi i\, a_A(z) } \, T_A\, \get^\ga,
\\[3mm] 
\get^{\ga\, \dag}_{qA}(z) =  
\gf^\dag_q(z) \, T(z) T^\dag_A T\inv(z) \, {\get^\ga}^\dag 
= \gf^\dag_q(z) \, e^{\mbox{-}2\pi i\, a_A(z) } \, T_A^\dag\, {\get^\ga}^\dag, 
}
\labl{TorusExp}
}
using the notation 
\( 
a_A(z) = a^I(z) w_I(T_A),
\)
and that  $w_I(T_A^\dag) = - w_I(T_A)$ because of the
Hermitean conjugation properties of the algebra \eqref{Algebra}.  
For $(-)^\ga = +$ the states $\gch^\ga_{qA}(x)$ are left--handed
spinors in four dimensions. 
The factor $1/\sqrt 2$ takes into account that only the positive
chiral four dimensional spinors are independent, because the ten 
dimensional gaugino is Majorana.  The gaugino wave functions 
$\get^\ga_{qA}(z)$ are periodic up to conjugation with $T_i$, as 
follows from the definition of $T(z)$ in \eqref{WilsonMat}. The
completeness of these gaugino mode functions can be stated as 
\equ{
\frac 12 \sum_{\ga, q}
\get^\ga_{qA}(z) \,\get^{AA'}\,  \get^{\ga\;\dag}_{q A'}(z') = 
\gd( z - z' - \gG)\,  \Id_G \otimes \Id_S.
\label{DecompIdentity}
}
Here $\Id_S$ denotes the identity in spinor space, $\Id_G$ is the
identity in the adjoint representation of the gauge group
$\E{8}\times\E{8}'$ and $\get^{AA'}$ is the inverse Killing metric 
defined below \eqref{Algebra}. The factor $\frac 12$ takes care of the
double counting over charge conjugate states.

%
%
\section{$\boldsymbol{\E{8}}$ Weyl reflections and classification of
$\boldsymbol{\E{8}}$ shifts} 
\labl{sect:WeylReflect}

Many statements involving the gauge shift vectors in the main text of
this article have been made up to Weyl reflections and lattice shifts. 
In this appendix we
define equivalent $\E{8}$ shift vectors, and briefly describe the
actions of some generators of the Weyl group of $\E{8}$ on those
vectors.  Some of the material found in this appendix is taken from 
\cite{Casas:1989wu}. 

The $\E{8}$ roots are given as
the roots and the weights of one of the spinor representations of $SO(16)$:
\equ{
\label{e8roots}
\bigl(\ubar{\pm 1, \pm 1, ~0^6}\bigr), \quad 
  ~ \text{and} ~
\bigl(\pm \mbox{$\frac 12$}, \ldots, \pm \mbox{$\frac 12$}\bigr)
~\text{with number of minus signs even}.  
}
All permutations of the underlined components give rise to 
roots of $SO(16)$. Let $\gG_8$ denote the root lattice of $E_8$. 
For all vectors in the lattice $\gG_8$ the sum over the entries  is even, 
since this holds for the roots \eqref{e8roots} that span this lattice. 
For  more details we refer to \cite{pol_2,DiFrancesco:1997nk}. 
Since a gauge shift $v$ has to fulfill $3 v^I w_I  \equiv 0$ for all
roots $w$, it follows that $3 v \in  \gG_8$ as the $\E{8}$ root
lattice is self--dual. 

Two $\E{8}$ gauge shifts $v$ and $v'$ are said to
be equivalent, $v \simeq v'$, if  
\equ{
v' = v + u, ~ u \in \gG_8 
\quad \text{or} \quad
v' = W_{\ga}(v)  = v - (\ga, v) \ga.
}
where $W_\ga(v)$ is called the Weyl reflection in root $\ga$ of
$\E{8}$. 

In general, the order in which two Weyl reflections are preformed, is
relevant    
\equ{
W_{\ga\gb}(v) = W_\ga (W_\gb(v)) = 
v - (\ga, v) \ga - (\gb, v) \gb + (\ga,\gb)(\gb, v) \ga. 
}
However, if the $\E{8}$ roots $\ga, \gb$ are orthogonal $(\ga,\gb) =0$,
clearly, their Weyl reflections do commute. 

Next, we describe the action of some Weyl reflections in more detail. 
The Weyl reflections corresponding to the $SO(16)$ roots act as 
\equ{
\label{signsaway}
(~v_1, ~v_2, ~v_3, \ldots) \simeq 
W_{(~1, \pm 1, ~0^6)}(~v_1, ~v_2, ~v_3, \ldots) = 
(\mp v_2, \mp v_1, ~v_3, \ldots).
}
Hence we see that by interchanging two shift elements, or replacing two
shift elements by minus those elements equivalent shifts are
obtained. In particular, if a shift has at least one zero, the sign of
all other entries is irrelevant. 
The action of the spinorial root $\ga = \frac 12\bigl( ~1^8~ \bigr)$
reads 
\equ{
v \simeq W_\ga(v) = v - \frac 14 (\sum_{I=1}^8 v^I) \bigl( ~1^8 ~\bigr). 
\labl{spinorialRootReflect} 
}

Let the shift be of the form $v = \frac 13\bigl( \pm, \ldots, \pm \bigr)$, 
and $\gD =  \text{diag}(\pm, \ldots, \pm)$ be a diagonal
$8\!\times\!8$ matrix with entries $\gD_I = \pm 1$, such that 
the product of the entries of $\ga = 3 \gD v/2$ is positive. Only then
$\ga$ is a spinorial root of $\E{8}$, which can be used to Weyl
reflect in 
\equ{
v \simeq W_\ga(v) = 
\bigl( \Id - \mbox{$\frac 14$} (\tr\, \gD) \gD \bigr) v
= (2 v_1, 2 v_2, 0^6) \simeq (-v_1, -v_2, 0^6), 
\labl{8nonzero}
}
where we used that $3^2 (v, \gD v) = \tr\, \gD$ and 
chose $\gD_1 = \gD_2 = -1, \gD_{I\neq 1,2} = 1$. The
final equivalence is valid, since the two vectors always
differ by a $\SO{16}$ root $(\pm 1, \pm 1,
0^6)$, depending on the signs of $v_1$ and $v_2$.

Consider two $\E{8}$ roots of the form 
$\ga = (a, a)$ and $\gb = (a,-a)$, then it is immediate that they are
orthogonal. On a shift vector $v = (r,~ s)$ that is also split into
two 4 component vectors, $r$ and $s$, their composition acts as 
\equ{
(r,~s) \simeq \tW_a(r, ~s) = W_{\ga \gb}(v) = 
\bigl( r - 2(a,r)a,~ s - 2(a,s)a \bigr),
\labl{doubleaction}
}
such that these vectors do not mix. Next we discuss an interesting
application of this formula in the same spirit as \eqref{8nonzero} above. 
Let $v$ be a shift vector which has
four or more components not equal to zero (modulo integers). By 
interchanging the components of this vector, it can be brought to the
form $v = \bigl(r, ~s\bigr)$, with $r= \frac 13(\pm, \ldots,
\pm)$. Furthermore, let $\gd = \text{diag}(\pm, \ldots, \pm)$ be a diagonal
$4\!\times\!4$ matrix with entries $\gd_i = \pm 1$. By taking  
$a = 3 \gd r/2$, we find that $\ga$, $\gb$ are weights of the spinor
representation of $SO(16)$. Applying \eqref{doubleaction} and using 
that $3^2 (r,\gd r) = \tr\, \gd$ gives
\equ{
\bigl(r, ~s\bigr) \simeq 
\tW_a\bigl(r, ~s\bigr) = 
\Bigl( 
\bigl( \Id - \mbox{$\frac 12$} (\tr\, \gd) \gd\bigl) r,~ 
s - \mbox{$\frac 92$} (r,\gd s) \gd r 
\Bigr) =  
\Bigl(
0^{i-1}, 2 r_i, 0^{4-i},~ 
s - \mbox{$\frac 92$} (r,\gd s) \gd r 
\Bigr).
\labl{4nonzero}
}
In the final equivalence we took $\gd_i =-1, \gd_{k\neq i} =1$ for a
fixed $i = 1,\ldots 4$. 
These examples of equivalence of $\E{8}$ shift vectors can be used to
determine to which standard shift an arbitrary $\E{8} \times \E{8}'$
shift $(v, v')$, that satisfies the requirement of modular invariance
\eqref{eq:modinv}, is equivalent. 
A set of standard shifts in the two $\E{8}$ factors is given in table 
\ref{tab:z3models}. The results of the following analysis have
been summarized in table \ref{tab:identifyShifts}.

First, we bring the entries of $3v\in \gG_8$ to a standard form:  
If $3v$ has a half-integer entry, all its entries are half integer,
therefore by adding any spinorial root to $v$ all entries of $3v$ become
integer. Since $(2, 0^7)$ and its permutations are the sums of two
roots of $\SO{16}$, we infer that the integer valued entries of $3v$
can be restricted to $3v^I =  -2, \ldots, 3$. In fact, we
may even assume that no entry $3 v^I$ is equal 3: If there are two or
more entries equal to 3, then by adding the $\SO{16}$ root with 
$-1$ at two of these entries, they become zero. If there is just one
entry equal to 3, there is at least one other entry of $3v^I$ equal to
$\pm 1$, otherwise the sum of entries of $3v$ is not even, i.e.\ 
$3v \not\in \gG_8$. Again, by adding  an appropriate $\SO{16}$ root we
can make the  $3$ entry $0$, and turn the $\pm 1$  entry into 
$\pm 2$. (We have assumed that this procedure has been applied throughout
the paper to set all entries $3v^I \in \{ -2, -1, 0, 1, 2 \}$.)

Let us first consider the case, that there is at least one entry $3v^J = 0$. 
It follows from (\ref{signsaway}), that signs do not matter anymore;
we take them positive. As the sum of entries $3v^I$ is even, it
follows that the number of $1$'s is even. By a $\SO{16}$ root 
any pair of $2$'s can be mapped to $1$'s, therefore all the
entries $3v^J=0,1$ if the number of zeros is even, or with one additional
entry of $2$ if the number of zeros is odd. However, the number of
zeros determines the equivalent 
standard shift.  We find, that if the $\E{8}$ shift has $8, 7, 6, 5$, or $3$
zeros, it is equivalent to the $\E{8}$, $\SO{14}$, $\E{7}$, $\E{6}$, or
$\SU{9}$ shifts, respectively, using permutations. This leaves the
shifts with $4, 2, 1$ and $0$ zeros to be considered, which have to be
treated separately. If the shift $3v$ has $4$ zero entries, the shift can
be represented by $(-1, 1^3, 0^4)$, since the signs of the entries do
not matter. Applying \eqref{4nonzero} with $r =(-1,1^3)$, we
infer that it is equivalent to the $\SO{14}$ shift. Similarly, if $3v$ 
has $2$ zero entries, it can be brought to the form
$(-1, 1^5, 0, 0)$. Using \eqref{4nonzero} again shows, that this
corresponds to an $\E{6}$ shift. If $3v$ contains only $1$ zero, we can
put it into the form $3v = (1^6, -2, 0)$, which is, using 
\eqref{spinorialRootReflect}, equivalent to $3v = (0^6, -3, -1)$. Adding
the root $(0^6, 1, 1)$ to $v$ we find the $\SO{14}$ shift vector again. 
If the shift does not have any zeros, all the entries of $3v$ can be chosen to
be $\pm 1$. This again follows because the sum of all entries of $3v$
is even, hence the number of $\pm 2$ is even, so that by adding
appropriate $\SO{16}$ roots it can be put in this form. 
When the product of all these entries is positive, we can  
employ \eqref{8nonzero} to show that the shift is equivalent to 
the $\E{7}$ shift. For the negative case, we split the shift into two
4 component vectors, $r, s$ and use \eqref{4nonzero}, with $\gd$
chosen such that $(r, \gd s) = 0$, to conclude that the shift is 
equivalent to the $\SU{9}$ shift.

%
%
\bibliographystyle{bibstyle.bst}
{\small
\bibliography{anomcanc}
}

\end{document}